 \documentclass[final,5p,times,twocolumn,authoryear]{elsarticle}

\usepackage{amssymb}
\usepackage{natbib}
\newcommand{\sortas}[1]{}
\usepackage{url}

\usepackage{graphicx}
\usepackage{caption}
\usepackage{comment}
\usepackage{tabularx}
\usepackage{ulem}
\usepackage[usenames,dvipsnames]{xcolor}
\usepackage{multirow}
\usepackage{threeparttable}
\usepackage{booktabs} 
\usepackage{tabularx, booktabs}%
\usepackage{algorithm} 
\usepackage{algpseudocode} 
\usepackage{amsmath}

\journal{Neural Computing and Applications}

\begin{document}

\begin{frontmatter}



\title{Can Large Language Models Beat Wall Street? Unveiling the Potential of AI in Stock Selection}

\author[att,unipi]{George Fatouros}\ead{george@alpha-tensor.ai}
\author[att]{Kostas Metaxas}\ead{kostas@alpha-tensor.ai}
\author[innov]{John Soldatos}\ead{jsoldat@innov-acts.com}
\author[unipi]{Dimosthenis Kyriazis}\ead{dimos@unipi.gr}

\affiliation[att]{organization={Alpha Tensor Technologies Ltd.}, 
            addressline={15 Bowling Green Lane}, 
            city={London},
            postcode={100190}, 
            country={UK}}
\affiliation[unipi]{organization={Department of Digital Systems, University of Piraeus},
            addressline={Karaoli and Dimitriou 80}, 
            city={Piraeus},
            postcode={18534}, 
            country={Greece}}

\affiliation[innov]{organization={Innov-Acts Limited},
            addressline={Kolokotroni 6}, 
            city={Nicosia},
            postcode={1101}, 
            country={Cyprus}}


\begin{abstract}

This paper introduces MarketSenseAI, an innovative framework leveraging GPT-4's advanced reasoning for selecting stocks in financial markets. By integrating Chain of Thought and In-Context Learning, MarketSenseAI analyzes diverse data sources, including market trends, news, fundamentals, and macroeconomic factors, to emulate expert investment decision-making. The development, implementation, and validation of the framework are elaborately discussed, underscoring its capability to generate actionable and interpretable investment signals. A notable feature of this work is employing GPT-4 both as a predictive mechanism and signal evaluator, revealing the significant impact of the AI-generated explanations on signal accuracy, reliability and acceptance. Through empirical testing on the competitive S\&P 100 stocks over a 15-month period, MarketSenseAI demonstrated exceptional performance, delivering excess alpha of 10\% to 30\% and achieving a cumulative return of up to 72\% over the period, while maintaining a risk profile comparable to the broader market. Our findings highlight the transformative potential of Large Language Models in financial decision-making, marking a significant leap in integrating generative AI into financial analytics and investment strategies.
\end{abstract}




\end{frontmatter}




\section{Introduction}
\label{introduction}

\subsection{Background and motivation}
\label{sec:1}
\begin{figure*}[ht]
    \centering
        \includegraphics[width=\linewidth]{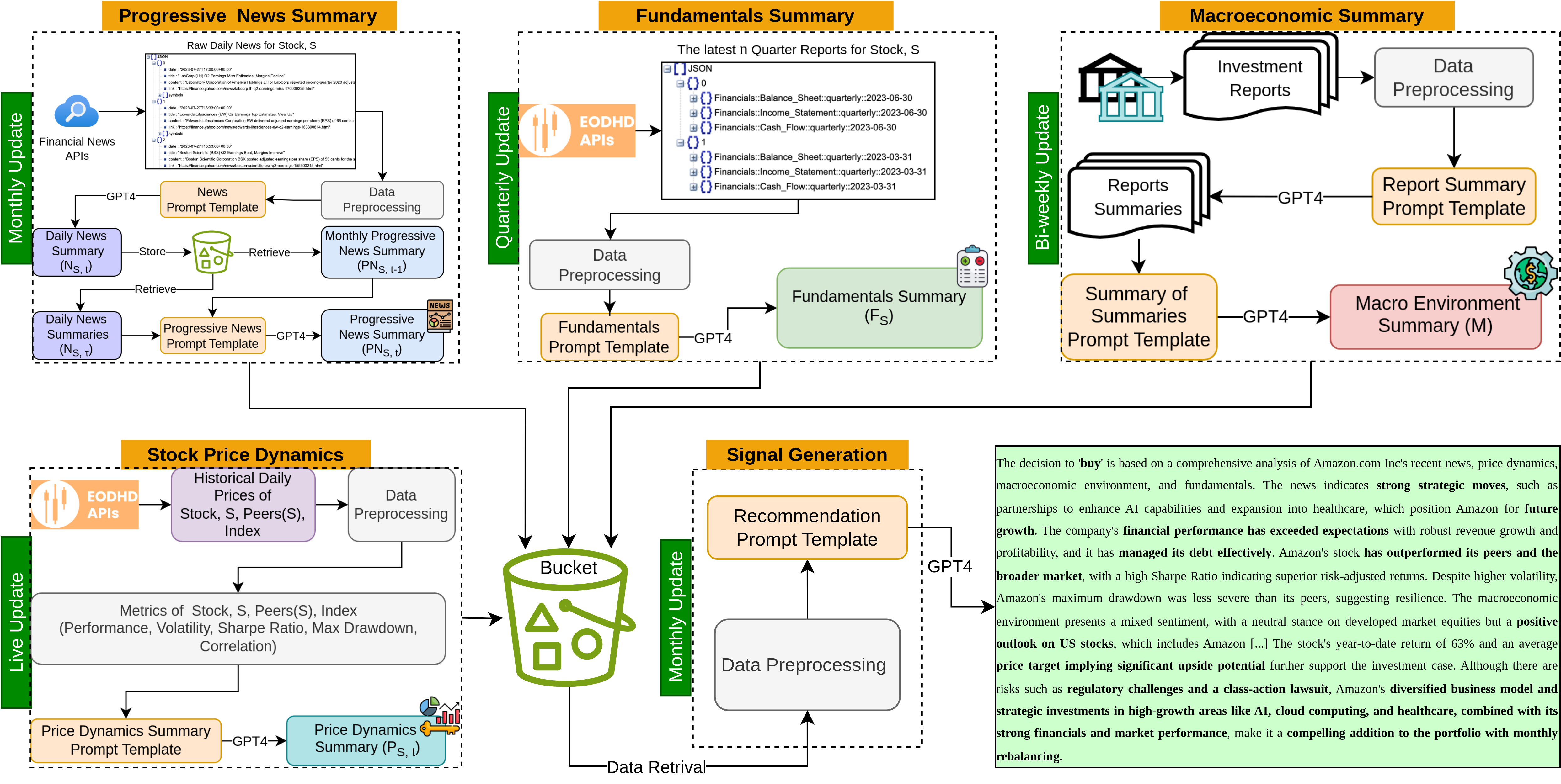}
        \caption{Conceptual architecture of MarketSenseAI, highlighting the core components, data flow, and outcome for a selected stock (e.g., Amazon).}
        \label{fig:marketsense}
\end{figure*}
Capital markets serve as an efficient conduit for capital allocation within an economy, and their price discovery process plays a key role in maintaining the health and stability of the financial system \cite{kidwell2016financial}. The price discovery process depends on a complex interplay of factors, including company and sector specific elements, macroeconomic data, momentum effects, as well as political and geopolitical influences \cite{lewellen2002momentum}. Market participants collectively participate in this intricate machinery of price discovery, thereby ensuring the efficient functioning of the financial markets \cite{malkiel2003efficient}.

Stock selection essentially operates as a price discovery mechanism through which market participants focus on stocks perceived as "mispriced", thereby offering the potential for attractive returns relative to the broader market. This principle forms the essence of value investing \cite{greenwald2020value}. However, the concept of "mispricing" can be expanded upon, as it may also relate to the market's perceived fair price of an asset, which may not necessarily align with its fundamental value. This can encompass expectations of future growth for a company, a strategy often referred to as "growth investing", which sometimes overlooks current fundamentals. Beyond the realms of value and growth investing, there are other factors that influence and add complexity to stock selection. The significance of passive investing, capital flows, derivative-related flows, and macroeconomic factors all contribute to a financial system that is inherently probabilistic and often chaotic by nature \cite{bouchaud2003fluctuations}. 

Market participants make decisions by understanding and using a wide range of information. However, retail investors often struggle with analyzing individual stocks. This challenge stems from their limited capacity to analyze information, susceptibility to behavioral biases, and lack of robust risk management skills. As a result, they might miss promising investment opportunities or expose themselves to undue risks. In this context, Exchange Traded Funds (ETFs) offer a practical solution. ETFs enable these investors to engage with the broader market more effectively, an approach often referred to as investing in "beta".

Similarly, small to medium-sized asset or wealth management firms encounter their own set of challenges. They may find it difficult to conduct in-depth analysis of individual stocks due to resource constraints or limited selection scope. For these firms, ETFs can also be an appealing option, providing a more manageable and diversified investment approach.

In contrast, larger professional firms are typically well-equipped with advanced technology, infrastructure, and skilled personnel. This enables them to conduct superior analysis and risk management of their investment portfolios. These firms often have dedicated teams of stock analysts, economists, and traders whose collective knowledge and reasoning ability are focused on capitalizing potential investment opportunities. However, even with these advantages, their ability to outperform the market is not guaranteed. They face unique challenges often associated with large-scale organizations, such as silos, poor communication, and diverse incentives \cite{weiss2019behavioral}.

During the last 15 years and following the financial crisis of 2008, there have been significant changes in the fabric and functioning of the capital markets, with lasting effects on price discovery. More specifically:
\begin{enumerate}
    \item Central banks policies: The 2008 crisis led to a strong belief among market participants that central banks would intervene to stabilize markets using every tool at their disposal. An over-reliance on central bank interventions, risks distorting market mechanisms and incentives as it can give rise to an underpricing of risk, and lead to moral hazard and potential increases in system-wide externalities \cite{bis-central-bank-tools}.
    \item The rise of passive investing: ETFs offer "blind" participation in market-weighted indices, treating all participating stocks irrespective of their fundamental value. This can allow stocks to substantially deviate from their fair value. It is evident that stocks that are widely held by passive investors are more likely to be affected than stocks that are not \cite{goyal2015passive}.
    \item The significant impact of retail investors: The emergence of retail investors, with easy access to gamified, leveraged, and derivative-enabled trading platforms, has also significantly impacted price discovery. One example is a highly popular retail product, zero-day expiry options (0DTE), which are option contracts that expire within a single day \cite{brogaard2023does}. In 2022 0DTE accounted for approximately 43\% of the total S\&P 500 options volume compared with just 6\% in 2017. Another example is meme stocks, such as GameStop, where retail herd behavior drove the price of the stock to astronomical levels \cite{anand2022role}.
\end{enumerate} 

The aforementioned factors collectively disrupt the proper functioning of price discovery, leading to reduced incentives for investors to accurately assess risk and value assets. In today's market environment, this underscores the urgent need for more sophisticated tools. These tools should not only enhance the analytical capabilities of human decision-makers but also augment their capacity to navigate the increasingly complex and data-rich financial landscape. With such tools, investors can achieve precision and insight, critical for making informed decisions in the current market dynamics.

\subsection{Potential of LLMs in Stock Selection and Financial Analysis}

The emergence of Large Language Models (LLMs) like ChatGPT holds promise for substantial enhancements in financial analysis and stock selection. These sophisticated Artificial Intelligence (AI) systems, trained in vast amounts and types of corpora, have not only demonstrated the capacity to replicate intricate facets of human cognition but, in numerous instances, have surpassed them \cite{openai2023gpt4}.

Quickly parsing through vast financial data, LLMs can discern intricate details from earnings reports to macroeconomic studies and process vast amounts of unstructured data, such as news articles or expert opinions, more efficiently than human analysts \cite{guo2023close}. This swift deep content analysis enables them to pinpoint patterns often missed in traditional analysis \cite{alshami2023harnessing}.

While humans may be influenced by cognitive biases, LLMs provide a more objective lens. They operate largely free from the emotional and cognitive biases that can cloud human judgment, although some biases from their training persist \cite{abramski2023cognitive,atreidescognitive}. Furthermore, LLMs transcend the limitations of individual or team analysts, seamlessly scaling this capacity across products, markets and most importantly investors.

In addition, LLMs play a crucial role in minimizing biases in stock selection. Unlike human analysts, LLMs are not influenced by emotional or cognitive biases, providing a more objective perspective in financial analysis \cite{tjuatja2023llms}. This objectivity is critical in making unbiased investment decisions, as LLMs rely on data-driven insights rather than subjective judgments. While some biases inherent in their training data can persist, LLMs significantly reduce the influence of human biases, such as overconfidence or confirmation bias, on investment decisions \cite{abramski2023cognitive, atreidescognitive}. Furthermore, LLMs can process and analyze vast amounts of financial data, transcending the limitations of individuals or team analysts.

Although such AI systems may outperform humans in some specific tasks, their predominant value proposition lies in supporting human capabilities. They serve as robust tools that enhance decision-making, elevate the quality of analytical endeavors, and augment overall productivity \cite{noy2023experimental}. For instance, complex tasks such as consolidating multiple financial statements from diverse subsidiaries of a large corporation can be streamlined by an LLM-based system \cite{kim2023bloated}. The latter is capable of highlighting discrepancies, flagging outliers, and providing an executive summary, a task that would be both time-consuming and susceptible to errors if done manually.

Validating the aforementioned observations, prominent players in the financial sector, including JPMorgan and Bloomberg, have recently launched AI-powered initiatives. Announcements of an AI-enable advisory platform \cite{jpmorgan-gpt} and the release of a finance-centric LLM \cite{bloomberg-gpt} from these entities, respectively, reaffirm the prominence of generative AI within the financial sector. Furthermore, Morgan Stanley has utilized OpenAI's models to develop a chatbot that aids financial advisors by leveraging the bank's extensive research data \cite{ms-openai}. In a similar vein, Broadridge, through its subsidiary LTX, has introduced BondGPT, a chatbot powered by GPT-4 designed to assist institutional investors in bond trading \cite{bondgpt}. It is crucial to note, however, that many major financial institutions like Goldman Sachs and BlackRock have dedicated AI departments working on specialized initiatives. Nonetheless, the details of such innovations in investment banking are often closely guarded within these institutions for competitive and proprietary reasons, thereby limiting the public disclosure of comprehensive information about these significant developments.

\subsection{Paper Contributions}

This paper makes several key contributions to the evolving discourse on the fusion of AI and financial analysis, primarily through the presentation of an innovative service for stock analysis, named MarketSenseAI\footnote{MarketSenseAI is available at \url{https://www.marketsense-ai.com/}.}, rooted in the power of LLMs. The salient contributions are outlined as follows:

\begin{enumerate}
    \item \textbf{A novel LLM-driven Investment Service}: Integrating various data sources to provide holistic stock investment insights.
    \item \textbf{Explainable Investment Signals}: Explainable investment insights to empower investors and ensure transparency.
    \item \textbf{Versatile Use Cases}: MarketSenseAI's design allows for individual usage of service components, catering to diverse investor needs.
    \item \textbf{Empirical Evaluation}: Demonstrating the reliability and statistical significance of the service's recommendations.
    \item \textbf{Superior Performance}: Highlighting its potential to outshine high-performing indices.
    \item \textbf{An Independent Financial Advisor}: Democratizing access to premium investment insights for retail investors, asset managers and other stakeholders.
\end{enumerate}

In essence, this paper pioneers the integration of multi-source data analysis with the cognitive capabilities of LLMs to redefine stock selection and portfolio management. The resultant service not only enhances the quality of stock recommendations but ensures they are backed by robust, explainable reasoning. At the core of MarketSenseAI, the LLM generates concise summaries from vast amounts of numerical and textual data, extracting crucial insights about a company's developments and stock potential. Subsequently, it analyzes these summaries, considering the investment horizon, to make investment suggestions on specific stocks. This dual-process approach, harnessing the summarization and analytical power of AI, offers a sophisticated tool for investors navigating the complexities of the stock market.

Furthermore, MarketSenseAI's modular architecture allows for diverse applications in the financial domain. Each component of this architecture provides specific insights, such as news, fundamentals, and macroeconomic summaries, that can be exploited separately. It can facilitate the construction of AI-based portfolios using the generated signals and their explanations, offering a revolutionary approach to asset management. This framework can be tailored to make personalized investment decisions, taking into account user preferences on risk, investment horizon, and goal. This adaptability demonstrates the framework's potential in various areas of finance, extending far beyond stock selection.

Overall, MarketSenseAI bridges domain knowledge and the latest AI advancements, providing a novel and applicable system in investment finance. This tool holds significant promise for asset managers and retail investors seeking advanced financial advice, especially those with limited resources and access to premium financial services.

The rest of this paper is structured as follows: Section \ref{sec:2} reviews related works, particularly focusing on AI in investment finance and the significance of LLMs. Section \ref{sec:3} introduces MarketSenseAI, detailing its architecture and key components. Section \ref{sec:4} describes the evaluation methodology, touching upon the data utilized and the comparison methods. Empirical findings are shared in Section \ref{sec:5}. Section \ref{sec:6} concludes the paper summarizing the primary contributions and insights.

\section{Related Work}
\label{sec:2}

The integration of AI techniques in finance has experienced a pronounced surge in the past decade, increasingly overshadowing traditional statistical or algorithmic methodologies \cite{oecd2021}. Various dimensions of the financial domain have seen the infusion of AI, encompassing risk assessment \cite{fatouros2023deepvar}, banking services \cite{kotios2022deep}, and trading operations \cite{bloomberg2019algo}. Concurrently, the ascendance of foundational models, particularly the successive iterations of the Generative Pre-trained Transformer (GPT) and their affiliated chat interfaces, is effecting transformative shifts across diverse sectors. Notably, the financial sector is increasingly incorporating insights from such models into its business operations \cite{mckinsey2023}.

Despite the relatively recent public accessibility of these models — with ChatGPT API (Application Programming Interface), for instance, becoming openly available only in March 2023 — a plethora of research endeavors have surfaced, elucidating methodologies that harness these Generative AI frameworks to augment investment paradigms.

Research presented by \cite{zaremba2023chatgpt} elucidates ChatGPT's potential in finance, especially for tasks necessitating natural language processing capabilities, such as sentiment analysis of financial news and summarizing earnings reports. Such tasks demonstrate a marked correlation with stock market dynamics \cite{tetlock2008more}. Furthermore, \cite{lopez2023can} establishes the accuracy of ChatGPT in sentiment analysis of financial news, emphasizing the positive correlation between ChatGPT-generated scores and subsequent stock returns. Under several investment strategies, they found that ChatGPT outperforms traditional sentiment analysis methods, with the news sentiment contributing to impressive returns.

In comparative evaluations, \cite{li2023chatgpt} posits that ChatGPT and GPT-4 surpass domain-specific models like FinBERT \cite{araci2019finbert} and BloombergGPT \cite{bloomberg-gpt} in several tasks including named entity recognition and news classification. Notably, while FinBERT exhibits superiority in financial sentiment analysis over ChatGPT, the research lacks prompt engineering and utilizes a dataset that inherently favors FinBERT. In contrast, \cite{fatouros2023transforming} presents evidence that ChatGPT outperforms FinBERT in financial sentiment analysis both in terms of classification performance and correlation with actual returns, even when applied with zero-shot prompting. Zero-shot prompting enables ChatGPT to perform tasks without prior specific training which indicates that it can be effective in sentiment analysis based on its comprehensive training, despite no explicit financial data training.

Further, \cite{kim2023bloated} advocates the utility of GPT-3.5 in summarizing corporate disclosures, suggesting that sentiment derived from these summaries provides a more accurate predictor of stock market reactions than the original documents. This finding underscores the value GPT offers to investors seeking concise and targeted information. \cite{kirtac2024sentiment} extended this analysis, demonstrating GPT-3.5's effectiveness in sentiment analysis for predicting next-day stock returns, while outperforming FinBERT and lexicon-based models. \cite{yu2023temporal} and \cite{chen2023chatgpt} further explored sentiment indicators derived from LLMs, and how these can be used to enhance forecasting models for stock movements, indicating LLMs' superior performance in predicting stock directions.

In summary, current research on LLMs in financial applications aligns with and reinforces the methodologies underpinning each component of the proposed system. However, to the best of our knowledge, the presented approach is distinct both in its design and evaluation methodology as it leverages multi-modal financial data, instead of barely news or news with historical prices, to deliver actionable and interpretable investment recommendations for the analyzed stocks while it outperforms high-performing ETFs. Unlike traditional methods that lean heavily on quantitative analysis, where sentiment indicators are used as features of predictive models, MarketSenseAI emphasizes language understanding and reasoning to generate investment insights after processing numerical and text data. This approach allows for the provision of detailed, AI-generated explanations for each recommendation, enhancing the interpretability and trustworthiness of the investment decisions. What is more, the evaluation considers transaction costs and the number of trades, highlighting MarketSenseAI's applicability in real-world settings.

\section{Methods}
\label{sec:3}

MarketSenseAI's architectural framework, depicted in Figure~\ref{fig:marketsense}, merges four core components responsible for data inputs with a fifth component to facilitate the final recommendation (i.e., buy, hold, or sell). This component synthesizes all the information and provides a concise explanation for the respective decision. Each component is built upon OpenAI's API and employs the GPT-4 model \citep{openai2023gpt4}, utilizing zero-shot prompting and in-context learning to execute distinct tasks \citep{dong2022survey}.

The framework is designed to emulate the decision-making process of a professional investment team. This process involves tracking recent developments of the company or its sector (via a news summarizer), analyzing the company's latest financial statements (through a fundamentals summarizer), and conducting a macroeconomic analysis of the prevailing environment while taking into account the price action dynamics (via marco and price dynamics summaries). 

The following paragraphs present the key components of this architecture in detail.

\subsection{Progressive News Summarizer}
\label{sec:3.1}

The influence of company-specific news — encompassing announcements, reports, analyst opinions, and research findings — on market sentiment and subsequently stock prices cannot be understated \citep{malik2011estimating}. Depending on their content, such news could have a short, long-term, or minimal influence \citep{alqahtani2020impact}. Hence, the sourcing and interpretation of news, demand meticulous handling in stock analysis.

\begin{figure}[ht]
    \centering
        \includegraphics[width=\linewidth]{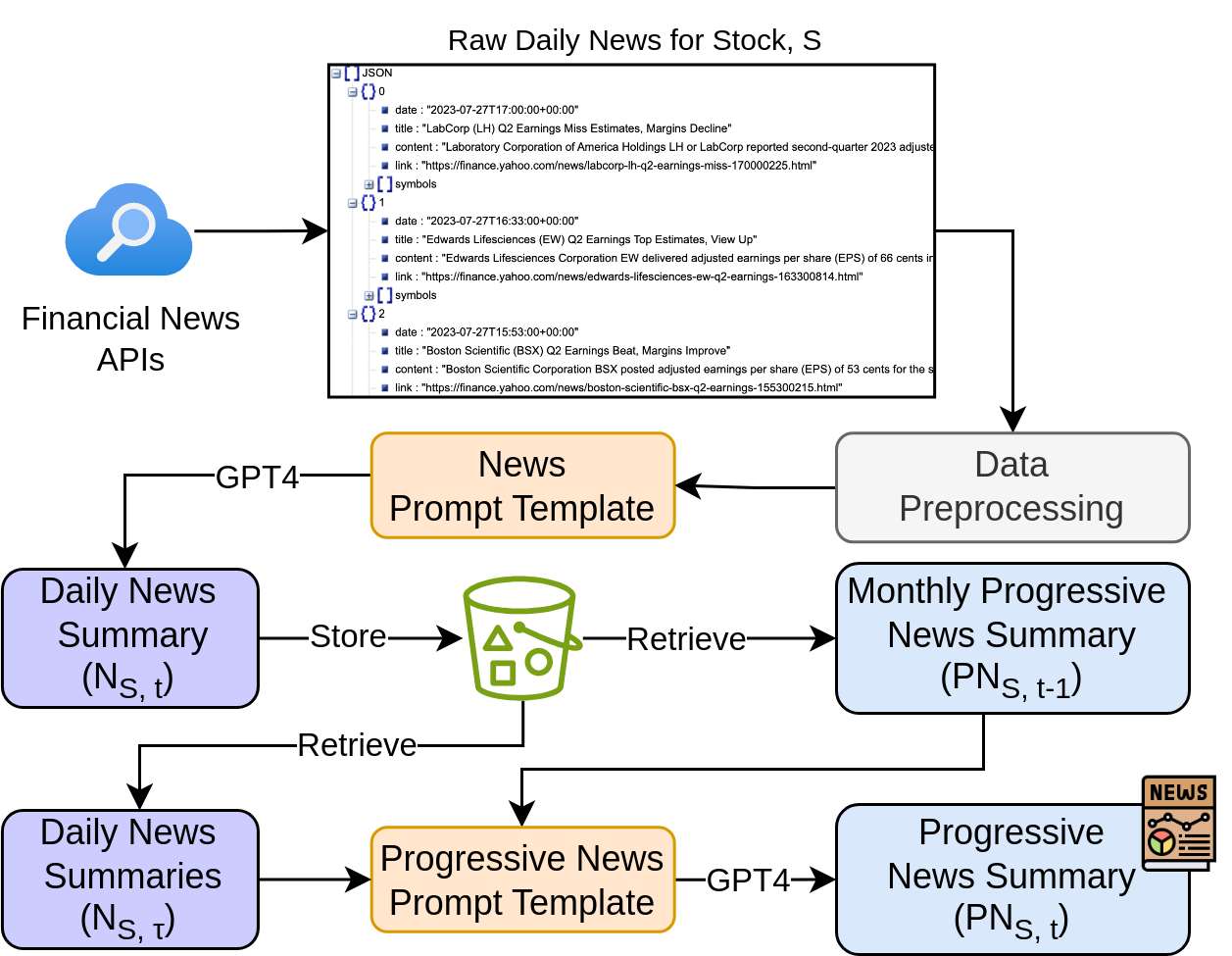}
        \caption{Progressive News Summarizer}
        \label{fig:news}
\end{figure}

The Progressive News Summarizer ($PN_{S,t}$) is responsible for news acquisition, condensation, and crafting a progressive synopsis of a stock's most influential news. As portrayed in Figure~\ref{fig:news}, daily news items corresponding to a specific stock are fetched from available APIs. The study utilized EODHD\footnote{\url{https://eodhd.com/financial-apis/stock-market-financial-news-api/}} Stock Market and Financial News API for news sourcing. Equation \ref{eq:news} and \ref{eq:prog_news} model the processes for obtaining the daily and the progressive news summaries, respectively.

\begin{equation}
\label{eq:news}
N_{S, \tau} = \bigoplus_{i=t-\tau}^{t} \text{Summarize}(N_{S, i})
\end{equation}

\begin{equation}
\label{eq:prog_news}
PN_{S,t} = \text{Summarize}(PN_{S,t=t-1}, N_{S, \tau})
\end{equation}

Where,

\begin{align}
    PN_{S,t} &: \text{ Progressive News Summary for stock } S \nonumber \\
    & \quad \text{ at time } t. \nonumber \\
    \text{Summarize}() &: \text{ Function to synthesize an updated summary.} \nonumber \\
    N_{S, \tau} &: \text{ Aggregated news summaries for stock } S \nonumber \\
    & \quad \text{ over the latest } \tau \text{ days.} \nonumber\\
    N_{S, i} &: \text{ Available news articles for stock } S \text{ on day } i. \nonumber \\
    \tau &: \text{ Number of days included in the aggregation, } \nonumber \\
    & \quad \text{ representing the time window for news } \nonumber\\
    & \quad \text{ progressive summary.}\nonumber\\
     \bigoplus &: \text{ Denotes the operation of concatenating } \nonumber \\
    & \quad \text{ daily news summaries.} \nonumber 
\end{align}

The daily news for a company is preprocessed to exclude text unrelated to the company, such as clickbait articles, thus ensuring it is in an appropriate format for entry into the prompt. GPT-4, accessed via OpenAI's API, is systematically prompted to distill the stock's daily news and generate a concise daily news summary ($N_{S, i}$). The latter is stored in a centralized repository.

While this method provides a summary for a specific date, it is essential to include the ongoing narrative of news-related content for the company, particularly older news that remains significant. For example, in the case of a merger or legal dispute, an announcement about the company's better-than-expected results may be less influential in the overall decision-making process. The Progressive News Summarizer addresses this by integrating the latest news summaries ($N_{S, \tau}$) with the preceding progressive summary ($PN_{S, t-1}$). More specifically, the prompt structure instructs GPT-4 to mimic the role of a financial analyst with the task of synthesizing an updated summary for a specific stock by integrating various types of information. The prompt includes:
\begin{itemize}
    \item \textbf{Current Summary}: The latest summary of the company and its stock as of a specific month and year.
    \item \textbf{Daily News Summary}: News articles, divided into factual news and analysts' opinions for the company over a given month.
    \item \textbf{Instructions}: Integrate the most pertinent information, distinguish factual news and analysts' opinions.
\end{itemize}

This summary is a synthesis that takes into account both the daily news summaries of the past month and the Progressive News Summary formulated in the previous month. This process ensures that the summarizer consistently reflects the latest, most relevant, and significant developments, offering a comprehensive and current snapshot of the company’s status in the news.

While the Progressive News Summarizer in this study is tailored for monthly intervals, its design offers adaptability for various frequencies to match different investment approaches. The effects of adjusting the summarizer's frequency for different intervals on the effectiveness of diverse investment strategies, remain an area for future exploration.

As illustrated by Table~\ref{tab:Apple}, the Progressive News Summarizer effectively captures the evolving narrative surrounding Apple Inc. (AAPL.US) over two distinct months in 2023. The summaries encompass a wide range of topics, from the company's ongoing financial performance to its strategic initiatives and market challenges.

In October, the focus was on Apple's significant role in the tech industry, marked by the launch of the iPhone 15 series and updates to its Apple Watch and AirPods. This period also highlighted challenges in smartphone sales, the impact of geopolitical issues, and fluctuations in stock performance. Unique to this month were reports on Apple's interest in acquiring Formula 1 broadcasting rights and the sale of company stock by CEO Tim Cook, underscoring the company's diverse strategic interests and executive decisions.

By November, while many earlier themes continued, new elements emerged. The summary shed light on Apple's sales slowdown, competitive pressures in the smartphone market, and its strategic shift in partnerships, including the termination of its credit card agreement with Goldman Sachs. Notably, the company's sustainability efforts and the launch of the M3 chip were highlighted, alongside its expansion in streaming content.

This table demonstrates the adaptive capability of the Progressive News Summarizer to integrate the latest corporate developments, market dynamics, and strategic maneuvers. It proves invaluable as a tool for delivering a comprehensive and current analysis of investment opportunities and industry trends. An example of a summary produced by this component is provided in Table~\ref{tab:news-aapl}.

\begin{table}[ht]
\caption{Apple Inc. Progressive News Summary (October vs November 2023)}
\label{tab:Apple} 
\small
\begin{tabular}{{p{4.5cm}p{1.5cm}p{1.5cm}}}
\hline
\textbf{Topic} & \textbf{October} & \textbf{November} \\
\hline
Financial Performance & Yes & Yes \\
Guidance for Q4 2023 & Yes & Yes \\
iPhone 15 Launch & Yes & Yes \\
New Apple Watch and AirPods & Yes & No \\
Smartphone Sales Challenges & Yes & Yes \\
Geopolitical Issues Impact & Yes & Yes \\
Stock Performance & Yes & Yes \\
Partnership with DuckDuckGo & Yes & No \\
CEO Tim Cook's Stock Sale & Yes & No \\
Interest in Formula 1 Broadcasting & Yes & No \\
Product Lineup including M3 Chip & No & Yes \\
Sales Slowdown and Competition & No & Yes \\
Appeal Win in UK & No & Yes \\
Partnership with Goldman Sachs & No & Yes \\
Regulatory Issues in Payment Apps & No & Yes \\
\hline
\end{tabular}
\end{table}

\begin{table*}[ht]
\centering
\caption{Apple Inc. Progressive News Summary (November 2023)} 
\label{tab:news-aapl}
\small
\begin{tabularx}{\textwidth}{lX}
\hline
\textbf{Category} & \textbf{News Summary} \\
\hline
Market Position & Dominant in the tech sector with record services revenue and robust product lineup, including the iPhone 15 series and new Mac products. \\
Sales Performance & Experiencing a slowdown, potentially dropping 5\% in iPhone sales due to challenges in China and Japan. \\
Stock Analysis & Recent dip viewed as a buying opportunity by analysts, seasonal performance aligns with product launch cycle. \\
Legal and Strategic Moves & Won UK appeal on mobile browser and cloud gaming services, potentially ending credit card partnership with Goldman Sachs. \\
Innovation and Sustainability & Launched M3 silicon chip based on 3-nanometer technology, expanding streaming content on Apple TV+. \\
Regulatory Challenges & Faces oversight of digital wallets and payment apps, navigating geopolitical and economic risks. \\
Investment Consideration & Despite challenges, presents a promising opportunity with strategic expansion, innovative products, and strong services division. \\
\hline
\end{tabularx}
\end{table*}

\subsection{Fundamentals Summarizer}
\label{sec:3.2}

\begin{figure}[ht]
    \centering
        \includegraphics[width=\linewidth]{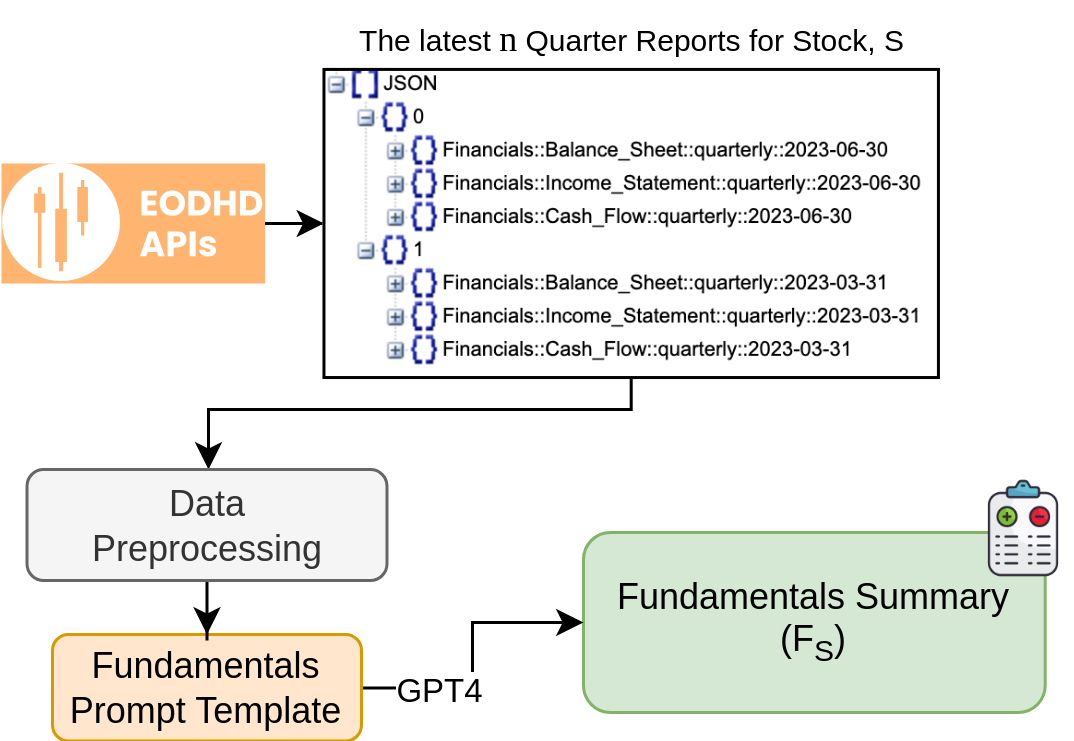}
        \caption{Stock's Fundamentals Summary}
        \label{fig:fundamentals}
\end{figure}

Fundamental data is crucial in predictive financial analytics, offering quantifiable metrics that reflect a company's current health and future trajectory. As depicted in Figure~\ref{fig:fundamentals}, we utilize EODHD's Fundamental Data API to source this quarterly information. To facilitate the comparison of financial data, we preprocess the data before inputting it into the prompt. This preprocessing includes a specific numerical abbreviation technique, which converts large numbers into a more compact format by representing them with prefixes like "million", "billion", or "thousand". For example, numbers in the billions are formatted as 'X billion', where X is the original number divided by 1 billion, rounded to two decimal places. This approach standardizes the financial data, ensuring consistency and clarity in the inputs fed to the LLM. We found that this preprocessing step is crucial for allowing GPT-4 to accurately compare and interpret complex financial figures.

Additionally, data from different quarters are placed side by side in a table format. The resulting prompt, fed to the GPT-4 model, delves into aspects such as profitability, revenue trajectory, debt metrics, and cash flow dynamics by comparing the most recent quarterly financial statements. This focus on recent data enables the LLM to detect shifts in financial performance, potentially correlating it with progressive news. More specifically, MarketSenseAI models the financial condition of company using the following formula:

\begin{equation}
    F_s = \text{Summarize}\left(\text{Standardize}\left(\bigcup_{i=1}^{n} \text{FinancialData}_{s, q_i}\right)\right)
\end{equation}

Where,

\begin{align}
    \text{FinancialData}_{s,q_i} &: \text{ Financial data for stock } s \text{ in quarter } q_i, \nonumber\\
    & \quad \text{where } i = 1, \dots, n \text{ (the last } n \text{ quarters).} \nonumber\\
    \text{Standardize}() &: \text{ Applies standardization techniques like } \nonumber \\
    & \quad \text{numerical abbreviation.} \nonumber \\
    \text{Summarize}() &: \text{ Generates a comprehensive summary } \nonumber \\
    & \quad \text{from this standardized data.} \nonumber
\end{align}

The prompt structure guides GPT-4 to adopt the role of a financial analyst focusing on recent trends. The AI is tasked with evaluating the financial health of a specified company's stock by analyzing its latest quarters. The prompt includes:
\begin{itemize}
\item \textbf{Financial Tables}: Key financial data from Balance Sheet, Income Statement, and Cash Flow for the latest quarters.
\item \textbf{Analysis Focus}: Recent trends and developments in profitability, revenue growth, debt levels, and cash flow generation.
\item \textbf{Instructions}: Conduct a bullet-form analysis.
\end{itemize}

Although LLMs traditionally face challenges in comprehending complex numerical data, our preprocessing approach, combined with GPT-4's capabilities, ensures accurate comparison and interpretation. The fundamentals summarizer is designed to present an unbiased, factual overview of a company's financial status, avoiding any direct investment recommendation. Table \ref{tab:fundamentals} showcases how the Fundamentals Summarizer distills key insights from financial statements, including income, balance, and cash flow statements. Similar to the Progressive News Summarizer, this component can be employed independently to provide a concise financial overview of the company being analyzed.

\begin{table*}[ht]
\caption{Apple Inc. Fundamentals Summary (2023-Q3)} 
\label{tab:fundamentals}
\small
\begin{tabularx}{\textwidth}{lX}
\hline
\textbf{Category} & \textbf{Fundamentals Summary}\\
\hline

Profitability & Apple's net income has increased from \$19.88 billion in Q2 2023 to \$22.96 billion in Q3 2023, indicating a strong profitability. The company's gross profit also increased from \$36.41 billion to \$40.43 billion over the same period.\\
Revenue Growth & Total revenue increased from \$81.80 billion in Q2 2023 to \$89.50 billion in Q3 2023, showing a positive revenue growth\\
Debt Levels & The company's total liabilities increased from \$274.76 billion in Q2 2023 to \$290.44 billion in Q3 2023. The long-term debt decreased slightly from \$98.07 billion to \$95.28 billion, but the short-term debt increased from \$11.21 billion to \$15.81 billion. The net debt also increased slightly from \$80.87 billion to \$81.12 billion. This indicates that the company's debt level is increasing, which could be a concern if not managed properly.\\
Cash Flow & The company's net cash from operating activities decreased from \$26.38 billion in Q2 2023 to \$21.60 billion in Q3 2023. However, the end period cash flow increased from \$29.90 billion to \$30.74 billion, indicating a positive cash flow generation.\\
Assets and Equity & Total assets increased from \$335.04 billion in Q2 2023 to \$352.58 billion in Q3 2023. The total stockholder equity also increased from \$60.27 billion to \$62.15 billion, indicating a growth in the company's assets and equity.\\
Conclusion & Apple Inc. shows strong profitability and revenue growth. However, the increasing debt level needs to be monitored. The company has a positive cash flow generation, and its assets and equity are growing.\\
\hline
\end{tabularx}
\end{table*}

\subsection{Stock Price Dynamics Summarizer}
\label{sec:3.3}

\begin{figure}[ht]
    \centering
        \includegraphics[width=\linewidth]{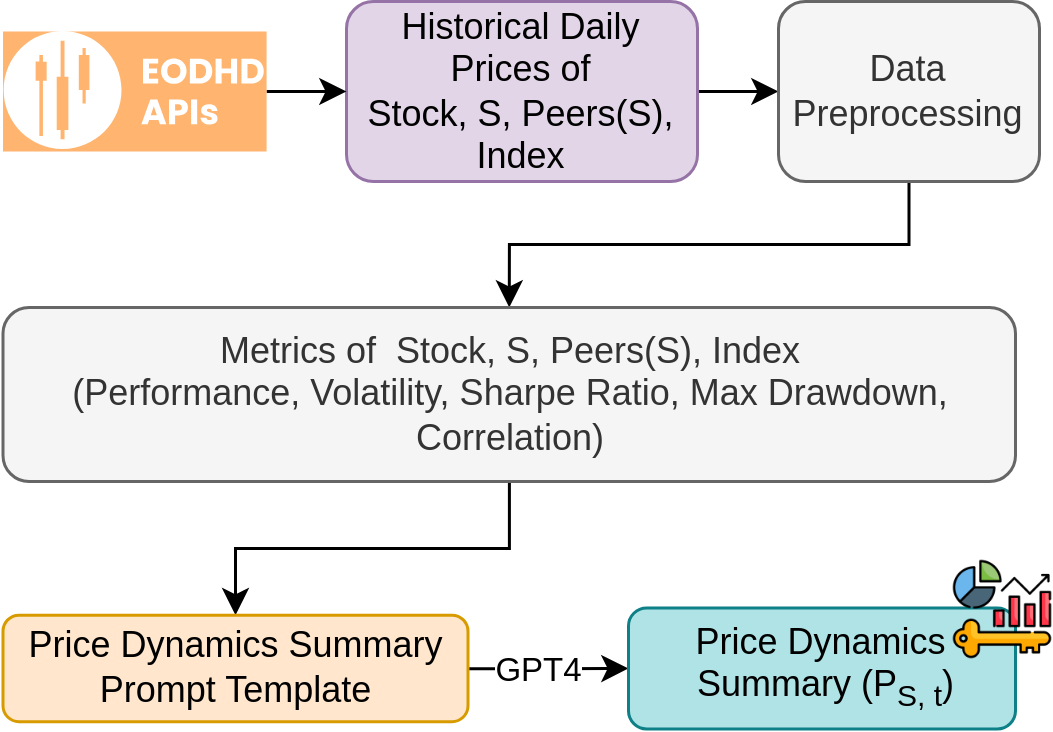}
        \caption{Stock Price Dynamics Summary}
        \label{fig:dynamcis}
\end{figure}

The Stock Price Dynamics Summarizer, a key component of MarketSenseAI, analyzes and contextualizes the price movements and financial metrics of stocks. As shown in Figure~\ref{fig:dynamcis}, this component not only examines the target stock but also compares its performance with the five most similar stocks, based on company description and sector, and includes the broader market context represented by the S\&P 500 index. The mathematical representation of the Stock Price Dynamics Summary is given by Equation~\ref{eq:priceDynamics}.

\begin{equation}
\label{eq:priceDynamics}
    P_{S,t} = \text{Summarize}\left(\text{Metrics}_{S, t}, \bigcup_{j=1}^{n} \text{Metrics}_{P_{j}, t}\right)
\end{equation}

Where,

\begin{align}
    P_{S,t} &: \text{ Stock Price Dynamics Summary }\nonumber \\
    & \quad \text{for stock  } S \text{ at time } t. \nonumber\\
    \text{Metrics}_{S, t} &: \text{ Performance metrics for stock } S \text{ at time } t. \nonumber \\
   \text{Metrics}_{P_{j}, t} &: \text{ Performance metrics for each peer stock } P_{j}  \nonumber \\
     & \quad \text{ at time } t.umber\\
    \bigcup &: \text{Union of metrics of the peer stocks for } \nonumber \\
     & \quad \text{comparison.}\nonumber\\
    \text{Summarize}() &: \text{ Function to condense the analysis } \nonumber\\
     & \quad \text{into a concise, factual report.}\nonumber
\end{align}

The prompt structure includes:

\begin{itemize}
\item \textbf{Performance Metrics}: Analysis of the stock's performance using metrics such as Cumulative Returns, Volatility, Sharpe Ratio, Maximum Drawdown, and a Correlation Matrix.
\item \textbf{Comparative Analysis}: Comparing the stock's performance with related stocks and the S\&P 500 index.
\item \textbf{Instructions}: Summarize the findings in a concise and factual report.
\end{itemize}

The methodology for identifying similar stocks is outlined in Algorithm~\ref{alg:similar_stocks}, utilizing the MPNet language model to generate embeddings and compute similarity scores \citep{song2020mpnet}. More specifically, stock descriptions are encoded by MPNet into high-dimensional vectors, capturing each company's unique characteristics and activities. This facilitates the computation of pairwise similarity scores among companies listed in the S\&P 500. These scores are crucial in identifying stocks with attributes similar to the target stock, thereby ensuring a comprehensive comparative analysis that integrates individual stock performance with broader market trends.

\begin{algorithm}
\caption{Identifying Stock Universe Using Embeddings}
\label{alg:similar_stocks}
\begin{algorithmic}[1] 
\Procedure{StockUniverse}{$targetStock, descriptions, n$}
    \State Generate embeddings for each stock description using a pre-trained language model
    \State Compute similarity scores between the target stock’s embedding and others
    \State Rank the stocks based on their similarity scores to the target stock
    \State Select the top $n$ stocks with the highest similarity scores
    \State \textbf{return} List of top $n$ similar stocks based on embeddings
\EndProcedure
\end{algorithmic}
\end{algorithm}

The summarizer fetches market data for the target stock, its similar stocks, and the S\&P 500 index. It analyzes key financial indicators, including cumulative returns and Sharpe ratios over 3, 6, and 12 months, and calculates volatility and maximum drawdown. These metrics, especially the Sharpe ratio and maximum drawdown, are crucial as they provide insights into the risk-adjusted returns and resilience of the stocks during market downturns, respectively \citep{korn2022drawdown}. This comprehensive analysis offers a broad understanding of the stock's performance relative to its peers and the broader market.

An example of the summarizer's output is illustrated in Table~\ref{tab:dynamics}, showcasing its ability to distill complex data into accessible insights. This provides a multi-dimensional view of the stock's market dynamics in relation to similar companies and the overall market trends.

\begin{table*}[ht]
\caption{Apple Inc. Stock Price Dynamics Summary (November 2023)}
\label{tab:dynamics}
\small
\begin{tabularx}{\textwidth}{lX}
\hline
\textbf{Metric} & \textbf{Price Dynamics Summary} \\
\hline
Cumulative Return & Apple Inc demonstrated a 29.0\% return, outperforming the S\&P 500 index's 13.7\% but underperforming tech peers like Adobe Systems and Amazon.com Inc. \\
Sharpe Ratio & Apple's Sharpe Ratio of 1.34 indicates a favorable risk-adjusted return compared to the market index Sharpe Ratio of 0.99, suggesting better compensation for the risk taken. \\
Volatility & Apple's volatility at 21.7\% is lower than that of Alphabet Inc, Adobe, and Amazon, indicating less erratic stock price movements. \\
Maximum Drawdown & Apple experienced a maximum drawdown of -16.0\%, which is less severe than the drawdowns of Adobe, Amazon, and Best Buy Co. Inc. \\
Correlation & Apple shows a high correlation with the S\&P 500 (0.76) and moderate correlation with other tech stocks like Microsoft Corporation and Amazon. \\
Conclusion & Apple has shown resilience and strong risk-adjusted performance relative to the broader market and some tech peers, with lower volatility and a relatively modest maximum drawdown. \\
\hline
\end{tabularx}
\end{table*}

\subsection{Macroeconomic Environment Summary} 
\label{sec:3.4}

Conducting an in-depth macroeconomic analysis is essential for making informed investment decisions and effective capital allocation. Such analysis offers vital insights into the overall economic health and performance, significantly influencing the profitability and value of individual companies as well as the broader stock market. By considering major forces that shape the investment landscape, such as the Covid-19 pandemic or the war in Ukraine, investors can make better-informed decisions.

\begin{figure}[ht]
\centering
\includegraphics[width=\linewidth]{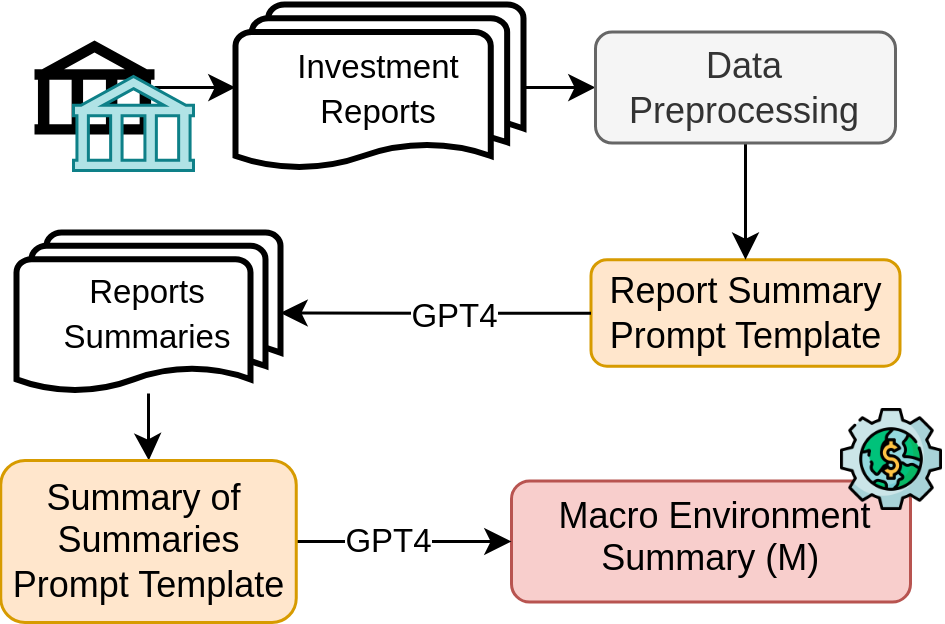}
\caption{Macroeconomic Environment Summary (MarketDigest)}
\label{fig:marketdigest}
\end{figure}

To facilitate this, MarketSenseAI includes a component named MarketDigest, depicted in Figure~\ref{fig:marketdigest}. This component synthesizes investment reports and research articles biweekly, providing succinct summaries of complex economic data and trends. MarketDigest sources information from a variety of publicly accessible reports from leading banks and investment institutions, including Goldman Sachs, Morgan Stanley, UBS, and BlackRock. The mathematical representation for MarketSenseAI's Macroeconomic Environment Summary component, MarketDigest, can be formulated as follows:

\begin{equation}
    M_t = \text{Summarize}\left( \bigcup_{j=1}^{N} \text{Summarize}\left(\text{Report}_{j,t}\right)\right)
\end{equation}

Where,

\begin{align}
    M_t &: \text{Macroeconomic summary at time } t.\nonumber \\
    \text{Summarize}() &: \text{Function that synthesizes data into a } \nonumber \\
     & \quad \text{concise overview.} \nonumber\\
    \bigcup &: \text{Union of individual summaries.} \nonumber \\
    \text{Report}_{j,t} &: \text{Investment report or article } j \text{ at time } t. \nonumber \\
    N &: \text{Number of reports/articles analyzed at time } t. \nonumber
\end{align}

The process begins by transforming these reports and articles into text form. Subsequently, the component utilizes GPT-4 to summarize individual reports, and then, in a second step, it condenses these individual summaries into a comprehensive overview. This approach allows MarketDigest to encapsulate diverse perspectives and analyses into a coherent narrative, offering a consensus view on the macroeconomic climate, central bank policies, preferred sectors or countries, as well as geopolitical trends. The output is concise yet thorough, accounting for potential contradictions or differing viewpoints among market analysts and experts. The prompt structure for MarketDigest includes:

\begin{itemize}
\item \textbf{Initial Summary Focus}: Summarization of individual reports, emphasizing critical macroeconomic elements including central bank policies, geopolitical insights, and market outlooks.
\item \textbf{Synthesis and Sentiment Analysis}: An in-depth analysis of all reports to extract consensus and divergent views, with focus on sentiment analysis categorized by asset class or investment dimension.
\item \textbf{Instructions}: Demanding a detailed and factual report, with emphasis on prevailing market sentiments and analytical categorization by asset class.
\end{itemize}

Table \ref{tab:macro} provides an example of a MarketDigest summary. Notably, KM Cube Asset Management\footnote{\url{https://www.km3am.com/}} initiated MarketDigest in March 2023 serving as a critical analytical tool, providing succinct market overviews to its clients. Distributed bi-weekly, it facilitates an enhanced understanding of market dynamics for both private and institutional clients, thereby informing their investment decision-making processes \citep{marketdigest-cube}. Furthermore, MarketDigest is integral to the company's monthly investment committee meetings, where it substantially contributes to the deliberation and formulation of strategies for investment portfolios and the management of discretionary products.

\begin{table*}[ht]
\caption{Macroeconomic Environment Summary (November 2023)} 
\label{tab:macro}
\small
\begin{tabularx}{\textwidth}{lX}
\hline
\textbf{Category} & \textbf{ Macro Summary} \\
\hline
Inflation & US core PCE inflation eased to 3.5\% in October, indicating a disinflation trend. \\
Interest Rates & Market-implied pricing suggests potential rate cuts in March 2024 for both the US and Europe. \\
Japan's Monetary Policy & Bank of Japan expected to weaken or abandon yield curve control due to domestic inflation. \\
ECB Policy & European Central Bank has begun balance sheet unwind. \\
Bond Market Outlook & Positive outlook on short- to medium-term developed market sovereign bonds. \\
Equity Market Stance & Neutral stance on developed market equities, with US stocks as largest allocation. \\
Global Growth & Global economy expected to experience below-trend growth in 2024. \\
Investment Strategy & Portfolios should maintain neutral exposure to risk and equities, overweight allocation to quality fixed income. \\
US Dollar & US dollar's position as leading global reserve currency showing signs of vulnerability. \\
Employment Risks & Risks to employment are on the downside, with leading indicators of employment deteriorating significantly. \\
Market Rally & Global financial markets experiencing significant rally, boosted by cooling inflation and falling Treasury yields. \\
Contradictions & Positive outlook on bonds but neutral on equities; US dollar vulnerability but remains a key currency. \\
Positive Sentiment & Short- to medium-term bonds, inflation-linked bonds, private market income, quality fixed income, US stocks. \\
Negative Sentiment & Credit, US Treasury, private markets, small-cap equities, Chinese equities. \\
Neutral Sentiment & Developed market equities, investment-grade credit, real estate, private equity funds, emerging markets outside China. \\
\hline
\end{tabularx}
\end{table*}

\subsection{Signal Generation}
\label{sec:3.5}

The signal generation component, as the final stage in the MarketSenseAI pipeline (Figure~\ref{fig:marketsense}), integrates the textual outputs from the news, fundamentals, price dynamics, and macroeconomic analysis components. This process results in a comprehensive investment recommendation for a specific stock, paired with a detailed rationale.

In essence, we argue that an investment decision for a stock is a function of the most important developments for the underlying company available in some extent in the news, the company's financial health, stock's performance in relation to competitors and the market, as well as the broader macroeconomic environment. The decision model is represented as: 

\begin{equation}
\label{eq:1}
    I_s = f(N_s, F_s, P_s, M)
\end{equation}

Where, $N_s, F_s, P_s$, and  $M$ are stock-specific ($s$) textual representations of current news, fundamentals, price dynamics, and macroeconomic conditions available from the components present in sections \ref{sec:3.1}-\ref{sec:3.4}.

We posit that the state-of-the-art LLM, GPT-4, possesses the requisite capacity to weigh and reason upon these different categories of data, as evidenced by its demonstrated proficiency in complex financial reasoning tasks \citep{callanan2023can}. In its operation, the GPT-4 model is prompted to adopt the role of an expert financial analyst. This approach employs a Chain of Thought methodology \citep{wei2022chain}, guiding the model through a logical, multi-step reasoning process that reflects an expert financial analyst's thinking pattern. By applying this technique, MarketSenseAI can effectively analyze and synthesize news, company fundamentals, stock performance data, and macroeconomic factors that could influence the given stock, thus providing reasoned and structured insights into stock selection. This approach is particularly useful in complex domains like finance, where the ability to navigate through multifaceted data and reason like an expert is crucial. Concurrently, in-context learning is employed to dynamically adjust the analysis based on current financial situations and evolving market data \citep{dong2022survey}. This dual strategy allows MarketSenseAI to provide deep insights that adapt in changing market conditions and investors' preferences, representing a significant advancement in AI-driven financial analysis. The prompt structure is as follows:

\begin{itemize}
    \item \textbf{News Analysis}: The model first evaluates the latest news summaries regarding the target company, assessing their potential impact on the stock's performance.

    \item \textbf{Price Dynamics Analysis}: Next, it compares the stock's price dynamics with those of related stocks and the overall market, offering a relative performance perspective.

    \item \textbf{Macroeconomic Environment Analysis}: The model then assesses the broader macroeconomic landscape and its implications for the target company, considering global economic trends and events.

    \item \textbf{Fundamentals Analysis}: The final step involves a review of the company's fundamental financial data, analyzing its current financial health and future prospects.

\end{itemize}

The model's output, structured in a concise format, includes a decision ("buy", "sell", or "hold") along with a clear, step-by-step explanation of the reasoning behind this choice. The terms "buy" and "sell" are defined within the context of portfolio positioning (long and short positions, respectively), while "hold" indicates no inclusion in the portfolio's composition regarding the specific stock.

Table~\ref{tab:ms-comparison} demonstrates MarketSenseAI's functionality in generating an explainable investment recommendation for Apple Inc. at the end of November 2023, suggesting a "hold" position in the rebalanced portfolio for the following month. The table reveals mixed signals, balancing positive developments such as strong financials and successful product launches against challenges like sales slowdowns, competitive pressures, and regulatory scrutiny. The overall decision to "hold" reflects a prudent approach, considering the various contrasting factors at play.

\begin{table*}[ht]
\centering
\caption{Apple Inc. Generated Signal and Explanation (November 2023)} 
\label{tab:ms-comparison}
\small
\begin{tabular}{p{1.5cm}p{16cm}}
\hline
\textbf{Decision} & \textbf{Explanation\tnote{a}} \\
\hline
HOLD  & 
The analysis [...] presents a mixed picture. Starting with the news analysis, while Apple remains a \textcolor{ForestGreen}{dominant player in the tech sector with solid financials and a successful product launch in South Korea,} it faces \textcolor{red}{challenges such as a sales slowdown, competitive pressures, and regulatory scrutiny. The Chinese government ban and sluggish sales in China and Japan are concerning,} \textcolor{ForestGreen}{but analysts see the recent stock dip as a buying opportunity, suggesting underlying strength}. The company's strategic moves, such as \textcolor{ForestGreen}{potentially ending its credit card partnership with Goldman Sachs, indicate a focus on its services business, which is a positive sign}. However, the \textcolor{red}{legal challenges and geopolitical risks cannot be ignored}. In terms of price dynamics, \textcolor{ForestGreen}{Apple has outperformed the S\&P 500 [...] strong risk-adjusted performance.} However, \textcolor{red}{the maximum drawdown and the high correlation with the market suggest that Apple is not immune to market downturns.} The macro environment analysis reveals a disinflation trend in the US and potential rate cuts in 2024, which could be \textcolor{ForestGreen}{favorable for equities}. However, the \textcolor{red}{global economy is expected to grow below trend}, and there is a neutral stance on developed market equities, suggesting caution. The \textcolor{ForestGreen}{positive outlook on quality fixed income and US stocks} is encouraging, but the \textcolor{red}{vulnerability of the US dollar and downside risks to employment add uncertainty}. Fundamentals analysis shows \textcolor{ForestGreen}{strong profitability and revenue growth for Apple, with increased net income and gross profit}. The company's \textcolor{red}{debt levels have risen}, which warrants monitoring, but the \textcolor{ForestGreen}{positive cash flow generation and growth in assets and equity are reassuring}. Considering the mixed signals [...] a "hold" decision is prudent[...]\\
\hline
\end{tabular}
\begin{tablenotes}
\small
      \item[a]{Note: Green text highlights positive developments for the company/stock, while red text marks elements with negative sentiment.}
\end{tablenotes}
\end{table*}

\begin{figure}[ht]
    \centering
        \includegraphics[width=\linewidth]{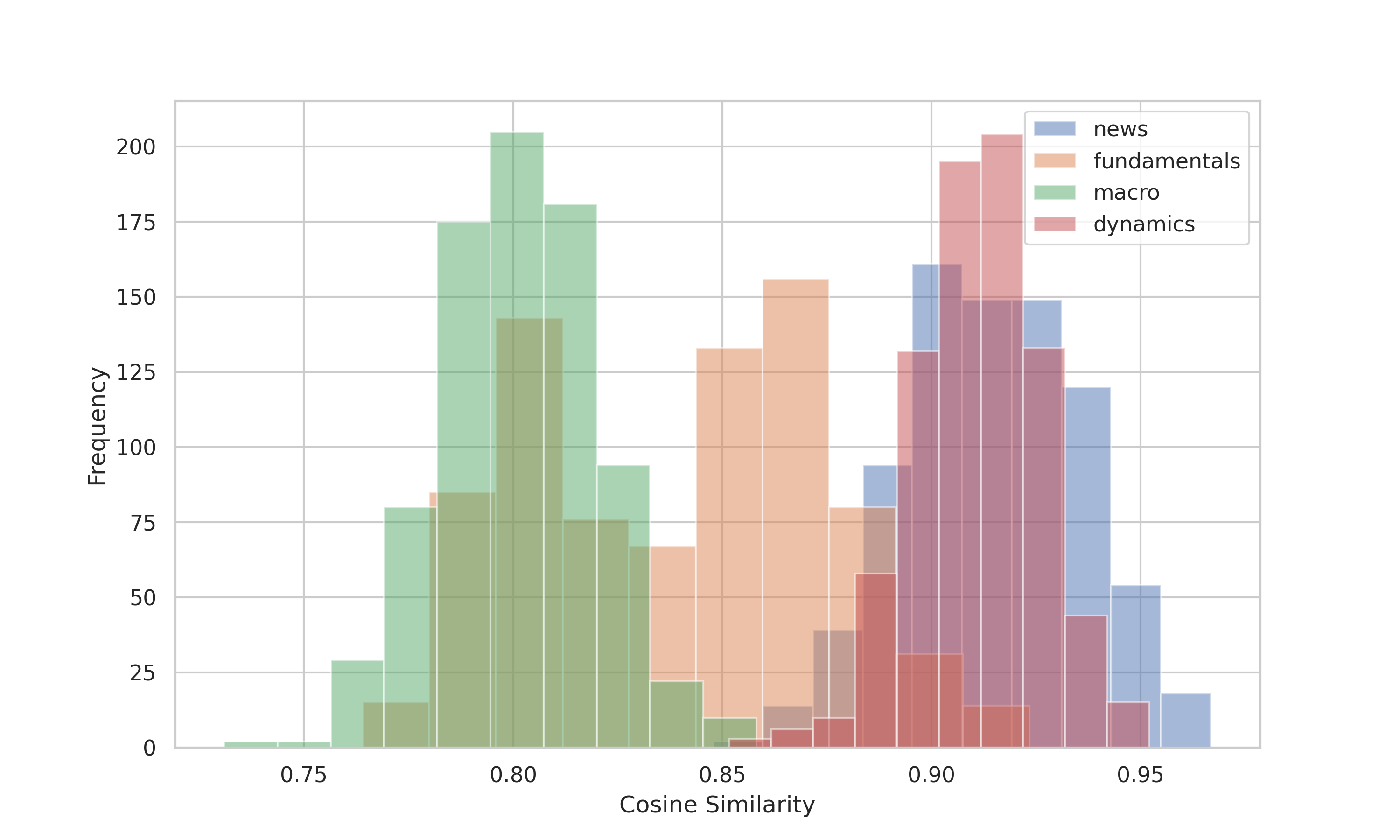}
        \caption{MarketSenseAI Components' Text Similarity with Signal}
        \label{fig:embeddings}
\end{figure}

The analysis of text similarity in MarketSenseAI, as shown in Figure~\ref{fig:embeddings} and detailed in Table~\ref{tab:embeddings}, reveals how the model's reasoning aligns with the dynamics of the input data, crucial for generating monthly investment signals. The high similarity scores in the "News" and "Price Dynamics" summaries (mean scores of 0.923 and 0.907, respectively) reflect the model's emphasis on these rapidly changing factors, recognizing their immediate impact on stock prices. This is especially relevant for short-term, monthly predictions where current developments and price trends can significantly sway market behavior.

In contrast, outputs from "Fundamentals" and "Macro" components, with mean similarity scores of 0.849 and 0.803, respectively, exhibit a lesser direct influence on the model's monthly decisions. The fundamentals, updated quarterly, provide a stable but less frequently changing view of a company's financial health, while macroeconomic data, being broader and more generic, have a more moderate impact on short-term investment decisions.

\begin{table}[ht]
\centering
\caption{Statistics of Text Similarity between Signals and Components  } 
\label{tab:embeddings}
\small
\begin{tabularx}{\linewidth}{lcccX}
\hline
\textbf{Component}     & \textbf{mean}  & \textbf{std}   & \textbf{min}  &  \textbf{max} \\
\hline
News  & 0.923573 & 0.016005 & 0.865069 &  0.967842 \\
Price Dynamics & 0.907652 & 0.013444 & 0.854822 &  0.942414 \\
Fundamentals   & 0.849269 & 0.033589 & 0.769349 &  0.934139 \\
Macro& 0.802891 & 0.017896 & 0.752009 &  0.870194 \\
\hline
\end{tabularx}
\end{table}

This analysis underscores MarketSenseAI's ability to process and integrate various data categories, tailoring its decision-making process to the nature of the input data. This approach is key to providing accurate and timely investment recommendations considering investment horizon.

\section{Experiments}
\label{sec:4}

This section outlines the data sources and methodologies employed in MarketSenseAI's analysis, along with the experimental setup used to evaluate and interpret the generated investment signals and explanations.

\subsection{Data}
\label{sec:4.1}
The evaluation of MarketSenseAI's performance leverages stocks listed in the S\&P 100 index, comprising the 100 largest and most established companies in U.S. equity markets. These stocks, due to their public visibility and the volume of analysis they attract, present a challenging environment for achieving superior stock-selection performance, especially when considering transaction costs \citep{invest2022}.

The assessment period spans from December 1, 2022, to March 31, 2024. During this time, MarketSenseAI utilized diverse datasets for its in-context learning processes:

\begin{enumerate}
    \item \textbf{News}: A total of 163,483 articles published from December 1, 2022, to February 29, 2024, averaging 4.57 articles per day per stock, with a standard deviation of 5.49. The mean number of tokens per article was 867, with a standard deviation of 1196. These data produced 35,229 daily, company-specific, news summaries and 1,500 monthly progressive summaries.
    \item \textbf{Fundamentals}: Financial data were gathered from 612 quarterly reports of the S\&P 100 stocks, starting from the second quarter of 2022. This data set produced 608 unique fundamentals summaries, averaging about 6 per stock.
    \item \textbf{Descriptions}: Concise descriptions of each stock and its sector, employed by Algorithm~\ref{alg:similar_stocks} to identify similar stocks.
    \item  \textbf{Prices}: Historical daily stock prices (adjusted close) from January 1, 2022, to February 29, 2024, were analyzed to compute stock price dynamics. This data was employed by the Stock Price Dynamics component to produce one summary per month for each stock, totaling 1,500 summaries.
    \item \textbf{Macro}: 187 investment reports (20-30 pages each) from major financial institutions published between April 2023 and February 2023, were analyzed by MarketDigest. For predictions made from January to March 2023, macroeconomic summaries were not available for the signal generation component, resulting in 11 macroeconomic summaries used for signal generation.
\end{enumerate} 

To evaluate the system, the "Signal Generation" component was fed at the end of each month with the latest available summaries (news, fundamentals, price dynamics, macro). While macroeconomic summaries were identical for all stocks, fundamental summaries were updated only when a new quarterly report for a stock became available. News and price summaries, being more dynamic, provided updated stock-specific insights each month. The investment signals generated by MarketSenseAI were subsequently assessed based on the actual stock performance in the month following the signal.  Overall, 1,500 signals produced (15 months x 100 stocks), a breakdown of these signals reveals 338 ”buy”, 1150 ”hold”, and 12 ”sell” signals.

\subsection{Evaluation}
\label{sec:4.2}

The evaluation of MarketSenseAI's generated signals focused on comparing them against bootstrapped signals and actual stock price movements under various investment strategies. Additionally, GPT-4's capability to rank "buy" signals based on their explanations was utilized as an indirect assessment of the quality of the signals and their accompanying explanations.

\subsubsection{Bootstrapping}
Bootstrapping is a statistical method that involves resampling data with replacement to estimate the variability of specific statistics, including standard errors, confidence intervals, and various accuracy metrics \citep{efron1986bootstrap}. This method is particularly useful when dealing with complex or unknown data distributions.

In this study, bootstrapping was employed to evaluate MarketSenseAI's performance and its statistical significance against randomized investment signals. For this purpose, a matrix of signals for MarketSenseAI ($[n_{months} \times n_{stocks}]$) was used, from which samples were drawn randomly, representing "sell" (-1), "hold" (0), and "buy" (1) positions. It is crucial to note that the distribution of these randomly generated signals might not mirror the distribution within the MarketSenseAI dataset.

This approach ensures that the findings are not due to random chance and provides a robust assessment of the model's performance. After an iterative examination, we settled on creating 10,000 random portfolios for bootstrapping, observing that additional samples did not significantly alter the evaluation outcomes.

\begin{figure}[ht]
    \centering
        \includegraphics[width=\linewidth]{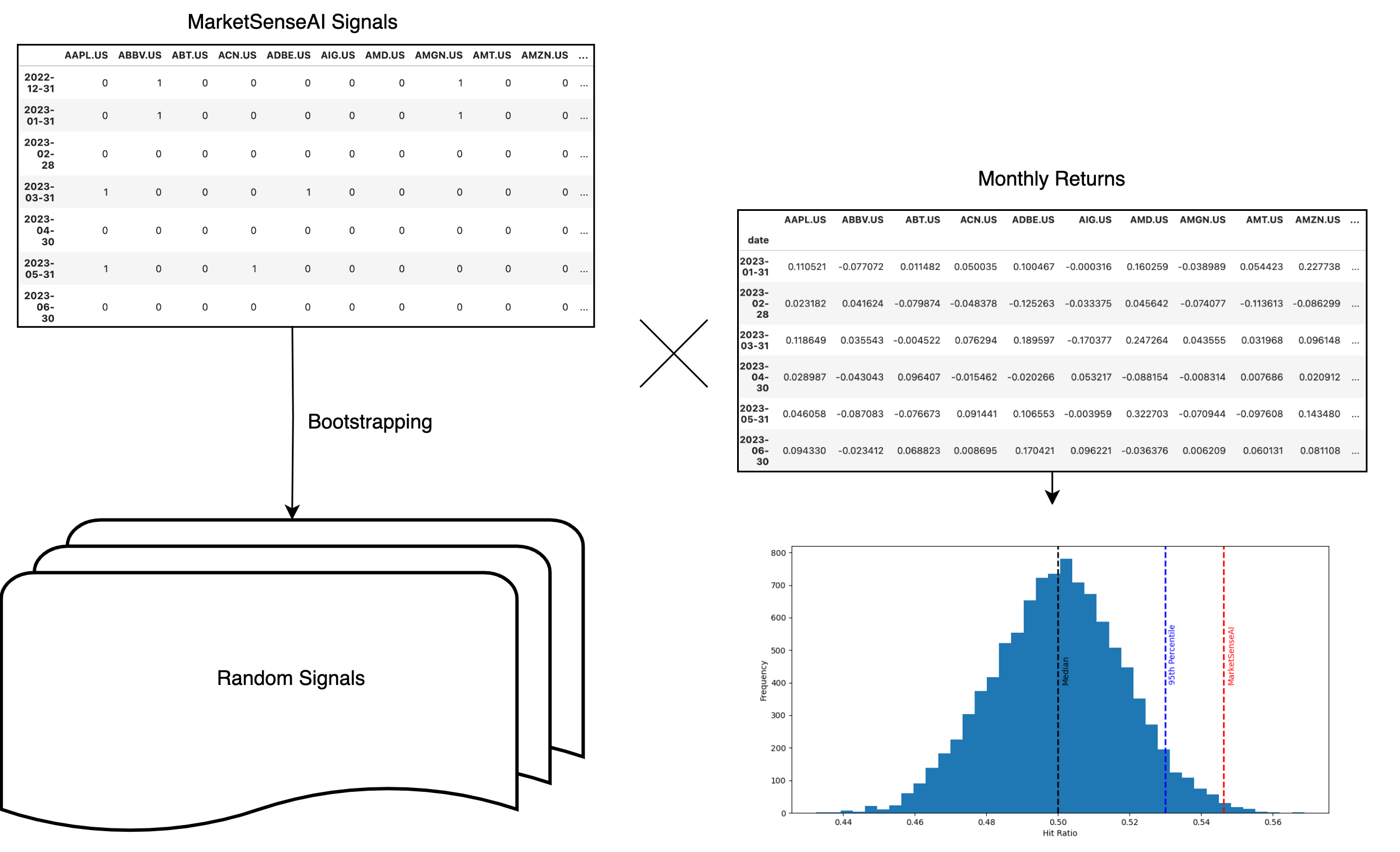}
        \caption{Bootstrapping-based Evaluation}
        \label{fig:bootstrapping}
\end{figure}

In this context, multiple randomized signals were generated for the stocks under consideration over the designated time frame. MarketSenseAI's performance was then compared against these randomized signals using two primary metrics. Firstly, the portfolio's cumulative returns were calculated by adhering to "buy", "sell", or both signals, applying equal weight to each and implementing monthly rebalancing (as depicted in Figure~\ref{fig:bootstrapping} and defined in Equation \ref{eq:2}). Secondly, the effectiveness of the signals was evaluated using a hit ratio, with the following month's actual returns serving as the reference benchmark (as detailed in Equation \ref{eq:3}).

The portfolio performance (cumulative return) is given by:
\begin{equation}
\label{eq:2}
    \text{Performance} = \prod_{i=1}^{n} \left(1 + \frac{\sum_{j=1}^{N} P_{L}(i, j)}{\text{signals per month at } i} \right)
\end{equation}

The hit ratio is calculated as:
\begin{equation}
\label{eq:3}
    HR_{L} = \frac{\sum_{(i, j) \in V_{L}} \text{I}(P_{L}(i, j) > 0)}{\text{length}(V_{L})}
\end{equation}

where,
\begin{align}
    P_{L}(i, j) &: \text{Performance of asset \( j \) at time \( i \), defined as } \nonumber \\
                & \quad\text{\( P_{L}(i, j) = m(i, j) \times r(i, j) \).} \nonumber\\
    L           &: \text{Indicator representing the evaluated signals, }L_{\text{long}}  \nonumber \\
                & \quad \text{ for long, } L_{\text{short}} \text{ for short, and } L_{\text{both}} \text{ for both signals.} \nonumber \\
    m(i, j) &: \text{ Model predictions (signals) for asset \( j \) at time \( i \).} \nonumber \\
    r(i, j) &: \text{ Actual returns for asset \( j \) at time \( i \).} \nonumber \\
    V_{L} &: \text{ Set of returns based on model predictions and \( L \).} \nonumber \\
    \text{I}(x) &: \text{ Indicator function, returning 1 if \( x \) is true} \nonumber\\
    & \quad\text{ and 0 otherwise.} \nonumber
\end{align}

This methodology ascertains the substantive impact of MarketSenseAI as it discerns the tangible benefits of MarketSenseAI's recommendations as opposed to adhering to random trading signals.

\subsubsection{Market Performance}
\begin{table}[h!]
\caption{Evaluated Investments Strategies on S\&P 100 Stocks}
\label{tab:portfolios}
\small
\begin{tabularx}{\linewidth}{lX}
\hline
\textbf{Abbreviation} & \textbf{Description} \\
\hline
MS & Equally weighted portfolio rebalanced monthly based on both "buy" and "sell" signals of MarketSenseAI.\\
MS-L & Equally weighted portfolio rebalanced monthly based on the "buy" signals of MarketSenseAI.\\
MS-L-Cap & Capitalization-weighted portfolio rebalanced monthly based on the "buy" signals of MarketSenseAI.\\
MS-Top10-SR & Equally weighted portfolio rebalanced monthly based on the 10 stocks with the best Sharpe Ratio of all the stocks with a "buy" signal.\\
S\&P100-Eq & Equally weighted portfolio of all the stocks of the S\&P 100 index.\\
S\&P100 & Capitalization-weighted S\&P 100 index (OEF ETF).\\
Naive & Equally weighted portfolio rebalanced monthly for all S\&P 100 stocks with price above their corresponding 200 day moving average, fully allocated. \\
Naive-Top10 & Equally weighted portfolio rebalanced monthly based on the 10 stocks with the best Sharpe Ratio and their price above their corresponding 200 day moving average, fully allocated.\\
MS-TopN-GPT & Equally weighted portfolio rebalanced monthly based on the $N$ stocks with the best score produced by GPT-4 after processing of all the stocks with a "buy" signal.\\
MS-High-GPT & Equally weighted portfolio rebalanced monthly based on stocks with score grater than 7/10 produced by GPT-4 after processing of all the stocks with a "buy" signal.\\
MS-Low-GPT & Equally weighted portfolio rebalanced monthly based on stocks with score lower or equal than 7/10 produced by GPT-4 after processing of all the stocks with a "buy" signal.\\
MS-TopN-Cap-GPT & Capitalization-weighted portfolio rebalanced monthly based on the $N$ stocks with the best score produced by GPT-4 after processing of all the stocks with a "buy" signal.\\
\hline
\end{tabularx}
\end{table}
This part of the assessment compares the performance of portfolios constructed based on MarketSenseAI's signals with actual market prices. The design of the MarketSenseAI-based portfolios and the baseline portfolios and the evaluation metrics are detailed in Table~\ref{tab:portfolios} and Table~\ref{tab:metrics}, respectively. The MarketSenseAI portfolios were formulated by following the service's signals, generated on the last day of each month after market closure, and held for one month.

\begin{table}[ht]
\caption{Portfolio Evaluation Metrics}
\label{tab:metrics}
\small
\begin{tabular}{p{2cm}p{6cm}}
\hline
\textbf{Metric} & \textbf{Description} \\
\hline
Total Return & The portfolio cumulative returns (\%) over a specific period (Equation \ref{eq:2}) \\
Sharpe Ratio & A measure of risk-adjusted return; calculated as the average return earned in excess of the risk-free rate per unit of volatility. \\
Sortino Ratio & Similar to the Sharpe Ratio, but measures returns relative to downside risk, focusing on negative asset volatility. \\
Volatility & A statistical measure of the dispersion of returns for a given security or market index, measured using standard deviation. \\
Win Rate & The percentage of trades that are profitable out of the total number executed. \\
Maximum Drawdown (Ddn) & The maximum observed percentage loss from a peak to a trough of a portfolio, before a new peak is attained. \\
\hline
\end{tabular}
\end{table}

\subsubsection{Ex-post evaluation with Ranking}

\begin{figure}[ht]
    \centering
        \includegraphics[width=\linewidth]{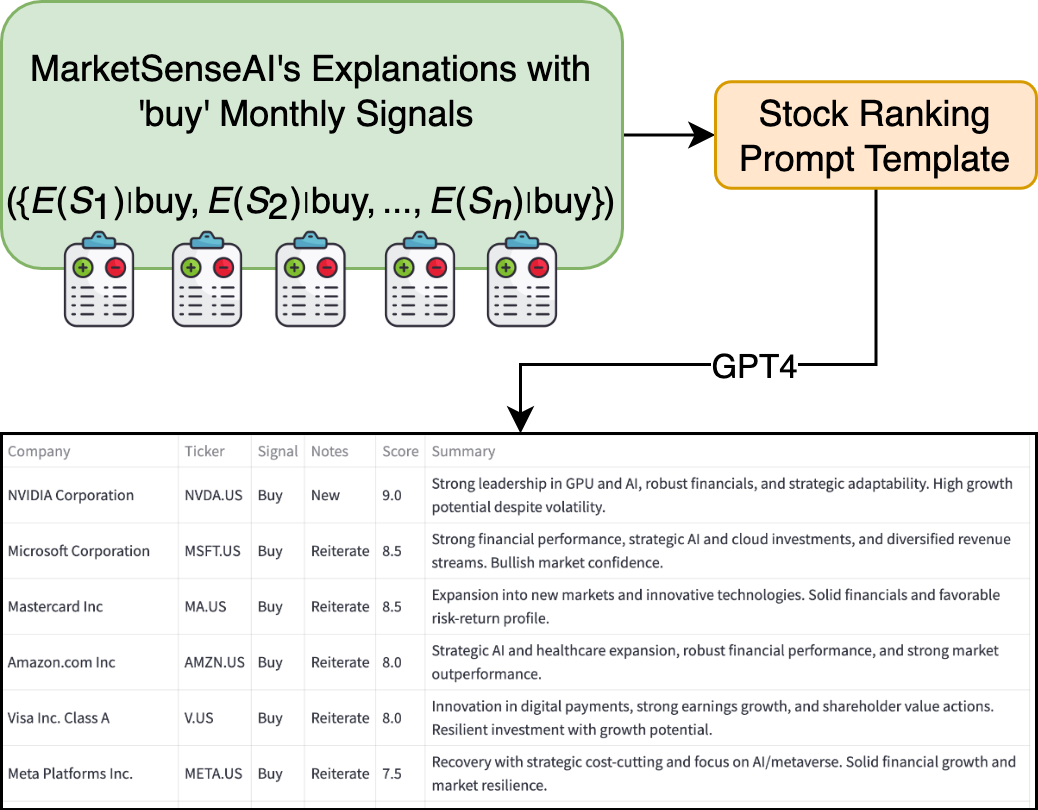}
        \caption{Signals Ranking by GPT-4}
        \label{fig:scoring2}
\end{figure}

An additional layer of analysis was introduced to evaluate the quality and context of the explanations accompanying the "buy" signals. This process involved incorporating a ranking mechanism into the system, as depicted in Figure~\ref{fig:scoring2}. The GPT-4 model was fed with all explanations that led to "buy" signals for each stock within a given month. The set included all explanations $E(S_i)$ associated with a "buy" indication for each stock $S_{i|S_1=buy}$. The prompt directed GPT-4 to rank these explanations on a scale of 0 to 10, with 10 indicating a strong buy.

The rationale behind this approach is that if MarketSenseAI's outputs are insightful, specific, and actionable, GPT-4 should be able to rank them effectively \citep{shu2023fusion, liu2023gpteval}. Stocks with higher scores are expected to contribute to better-performing portfolios. This method also provides an alternative ranking, filtering, and weighting mechanism for portfolio management. This evaluation approach was implemented in the last three investment strategies listed in Table~\ref{tab:portfolios}.

\subsection{Setup}

The implementation of MarketSenseAI was executed using Python 3.11, leveraging the LangChain framework \citep{lang2022} for prompt construction and utilizing OpenAI's API for accessing the GPT-4 model. Each component of MarketSenseAI, as outlined in Section~\ref{sec:3}, functions independently, running as a standalone script. The outputs from these components are systematically stored in a datastore, ensuring organized and efficient data management. As MarketSenseAI accesses GPT-4 through OpenAI's API, the operational aspects fall under OpenAI's operational jurisdiction. This arrangement allows our framework to leverage the capabilities of GPT-4 efficiently, ensuring consistent processing times per stock, while offloading the computational and hardware management responsibilities to OpenAI's robust infrastructure.

For the backtesting of the portfolios used in the evaluation process, the VectorBT PRO library\footnote{\url{https://vectorbt.pro/}} was chosen. This library is known for its versatility and efficiency in conducting financial analysis and backtesting investment strategies, making it a suitable choice for rigorously evaluating the performance of MarketSenseAI-generated portfolios.

To identify the most similar stocks of a given stock, a prerequisite for the Stock Price Dynamics Summarizer, we use the \texttt{mpnet-base-v2} model available from Hugging Face's Transformers framework and \texttt{cosine\_similarity} function from Scikit-Learn Python library.

Experiments are run on a desktop computer with an AMD Ryzen 5 5600x 6-Core CPU, 32GiB of RAM, and an NVIDIA GeForce RTX 3070 GPU. For data storage a dedicated cloud based S3 bucket was used hosted on Amazon Web Services (AWS).

\section{Results}
\label{sec:5}

This section presents the empirical findings obtained from the multifaceted evaluations outlined in Section \ref{sec:4}.

\subsection{Bootstrapping Evaluation Results}

Table \ref{tab:bootstrap} displays the outcomes of the bootstrapping evaluation, which is instrumental in contrasting the efficacy of MarketSenseAI with various bootstrapped portfolios. This evaluation includes an assessment of MarketSenseAI's performance with detrended returns, providing a refined analysis of its signal generation capability. The detrending of returns is mathematically expressed as:

\begin{equation}
\label{eq:detrend}
    r'(i, j) = r(i, j) - \overline{r(i, \cdot)}
\end{equation}
In this formula, \( r'(i, j) \) denotes the detrended return for asset  \( j \) at time \( i \), \( r(i, j) \) is the actual return, and \( \overline{r(i, \cdot)} \) is the average return at time \( i \) across all assets. This detrending process is crucial as it helps to isolate the performance of individual stocks from the broader market trends, thereby offering a clearer perspective on MarketSenseAI's signal precision.

The table evaluates both cumulative returns (\( R \)) and hit ratios (\( HR \)), along with their respective quantiles (\( Q_R \) and \( Q_{HR} \)), delivering an extensive view of the system’s effectiveness in comparison to randomized strategies.

\begin{table}[ht]
\centering
\caption{MarketSenseAI vs Bootstrapped Portfolios} 
\label{tab:bootstrap}
\begin{tabular}{lcccc}
\hline
\textbf{Signals} & $R$ & $Q_R$ & $HR$ & $Q_{HR}$\\
\hline
Detrend-Buy&  7.67 & 97.85 & 51.86 & 99.50 \\
Buy  & 35.48 & 98.70 & 60.34 & 99.35 \\
Detrend-Sell      & 22.62 & 100   & 72.73 & 100   \\
Sell& 0.10  & 100   & 63.64 & 100   \\
Detrend-Both      &  8.41 & 99.55 & 52.61 & 94.20 \\
Both& 33.87 & 100   & 60.46 & 100   \\
\hline
\end{tabular}
\begin{tablenotes}
\small
      \item[a]{Note: MarketSenseAI cumulative returns $R$ and hit ratio $HR$ (\%) and their quantiles ($Q_R$, $Q_{HR}$) in the bootstrapped distribution.}
\end{tablenotes}
\end{table}

The results from the bootstrapping evaluation reveal that MarketSenseAI’s signals notably outperform random chance, as evident from the high quantiles achieved in both cumulative returns (\( R \)) and hit ratios (\( HR \)) across diverse signal categories. This superior performance holds true even when assessing detrended returns, indicating MarketSenseAI’s proficiency in discerning profitable investment opportunities from broader market movements.

A particularly key observation is the high hit ratio quantile for the "Buy" signals following detrending. Considering the upward trend of the market during the evaluation period, this indicates that MarketSenseAI’s recommendations have a greater probability of success compared to a random signal generation approach. This finding is significant as it highlights the model's capability in effectively pinpointing potential market-outperforming opportunities.

In essence, the bootstrapping evaluation robustly demonstrates MarketSenseAI's capacity to produce trading signals that significantly surpass what would be expected by mere chance.

\subsection{Market Performance Evaluation Results}

It is important to note that the period of the experiments is characterized by disparate performances between the stocks. While technology giants and AI-centric companies experienced a robust year, others showed modest returns \citep{magnificent2023}. Our analysis places special emphasis on equal-weighted indices, which helps counterbalance this disparity and illustrates MarketSenseAI's potential.
Additionally all results presented below account for transaction costs to assess real-world applicability and effectiveness of MarketSenseAI-derived strategies.

\subsubsection{Vanilla strategies }
\begin{figure}[ht]
    \centering
        \includegraphics[width=\linewidth]{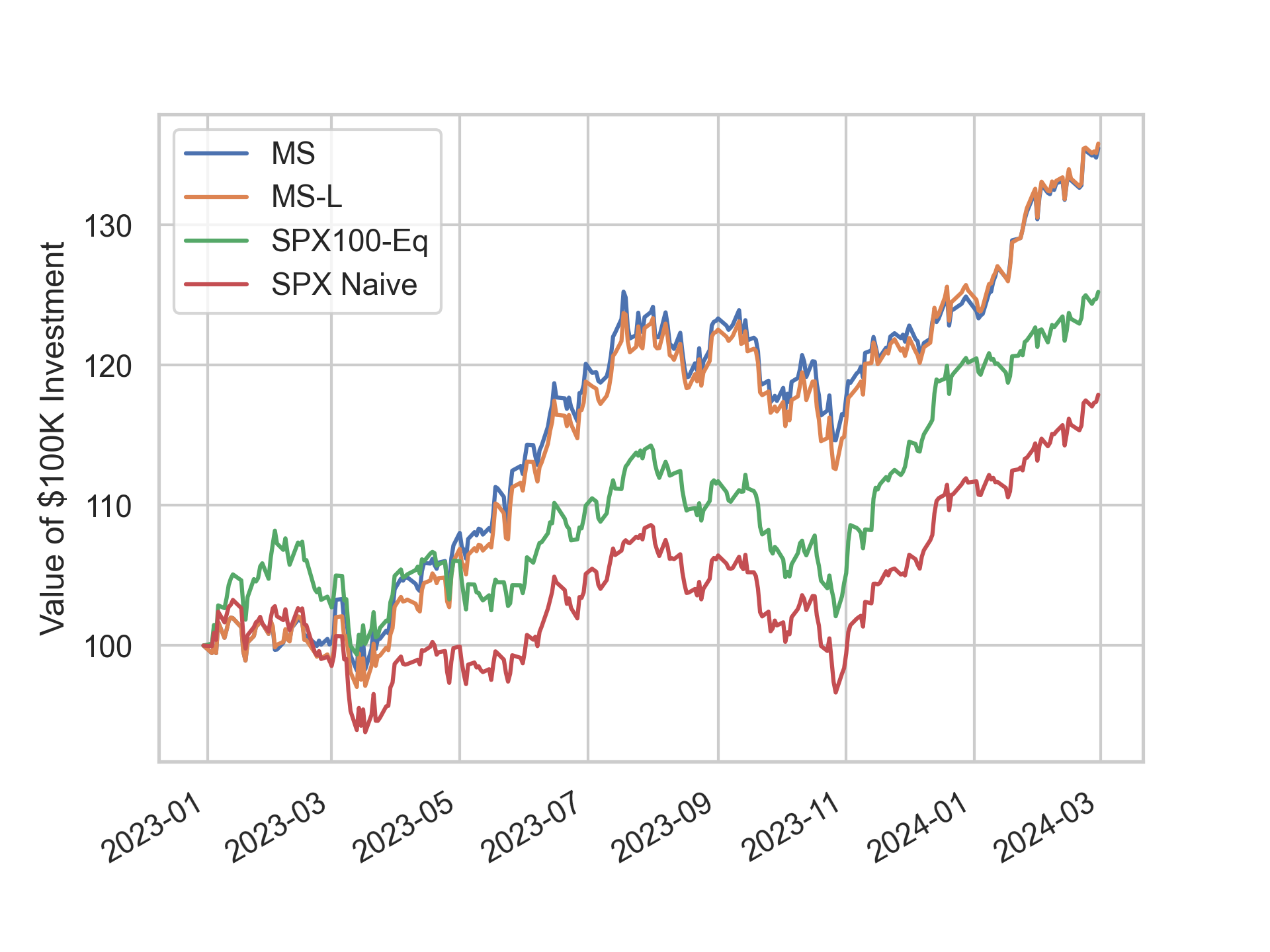}
        \caption{Performance of Equally-Weighted Portfolios}
        \label{fig:res1}
\end{figure}

The assessment of MarketSenseAI's vanilla strategies, as detailed in Table \ref{tab:res1} and illustrated in Figure \ref{fig:res1}, reveals the efficacy of LLM-driven investment strategies. The strategy following the complete set of MarketSenseAI's signals (MS) equally weighted, yields an impressive total return of 35.48\% (32.94\% after transaction costs) with a Sharpe and Sortino ratio of 2.49 and 3.87, respectively. The long-only version (MS-L) that takes into account only the "buy" signals of MarketSenseAI gives similar results given the relatively few "sell" signals generated. 

These results significantly outperform the equally-weighted S\&P 100 (S\&P100-Eq) both in total and risk-adjusted returns, with an "alpha" of approximately 10\% and 28\% higher Sortino ratio. The naive trend-following strategy (Naive), often used by market participants, yielded significantly lower results.

\begin{figure}[ht]
    \centering
        \includegraphics[width=\linewidth]{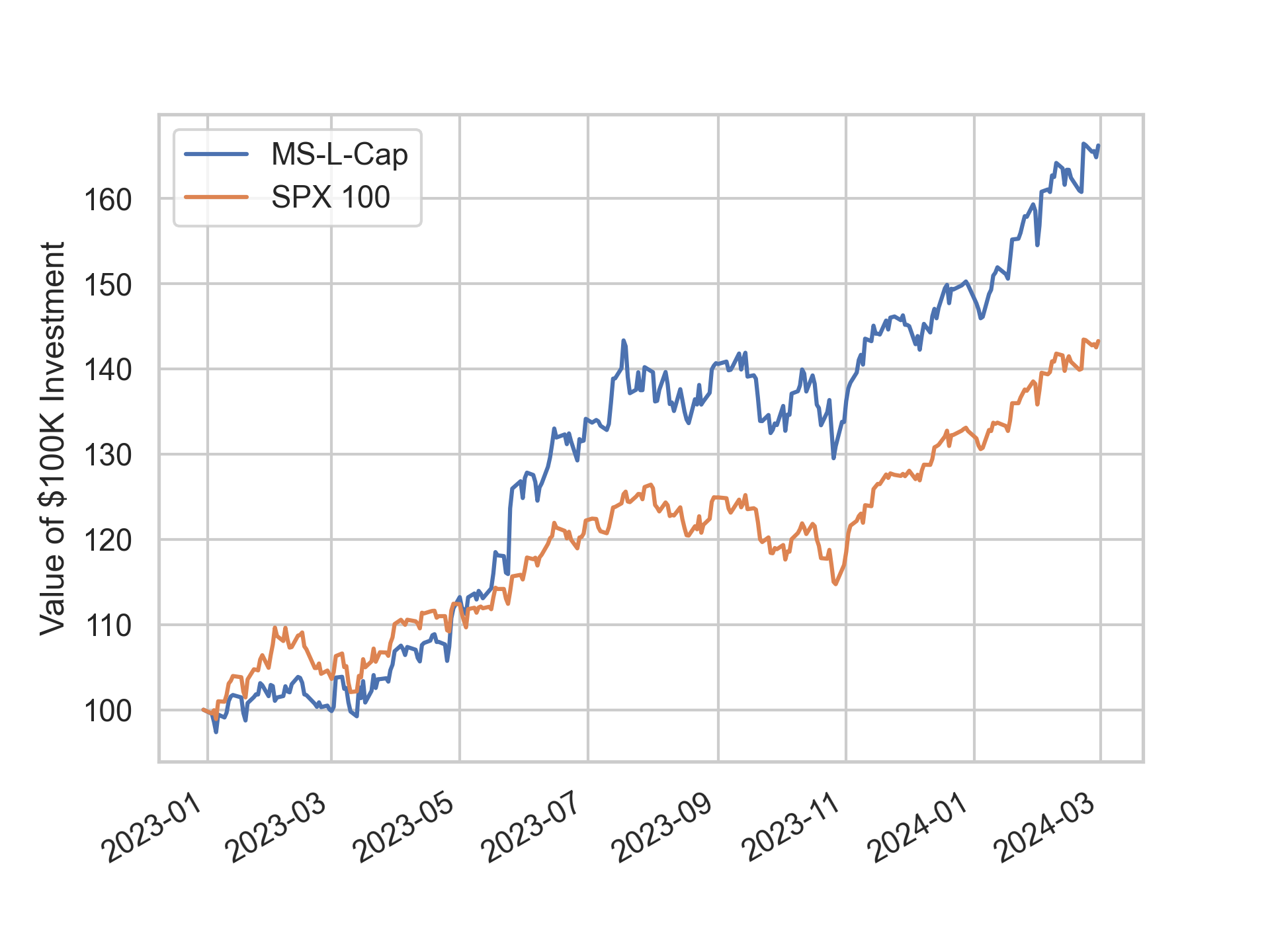}
        \caption{Performance of Capitalization-Weighted Portfolios}
        \label{fig:res2}
\end{figure}

In terms of capitalization-weighted performance, the MS-L-Cap emerges as a top performer in terms of Sharpe and Sortino ratios, as well as total return, reaching an impressive 66\% total return. This is a significant outperformance of 43\% to the S\&P100 ETF as depicted in Figure~\ref{fig:res2}.

\begin{table*}[ht]
\centering
\caption{MarketSenseAI Performance of Vanilla Strategies} 
\label{tab:res1}
\begin{tabular}{lcccccc}
\hline
\textbf{Strategy} & \textbf{Total Return(\%)} & \textbf{Sharpe} & \textbf{Sortino} & \textbf{Vol(\%)}  & \textbf{Win Rate(\%)} & \textbf{Max Ddn(\%)} \\
\hline
MS& 35.48 (32.94)& \underline{2.49} & \underline{3.87} & 15.76& 65.68 & \underline{8.47}\\
MS-L& \underline{35.79 (34.82)} & 2.41& 3.75& 16.44& 65.02& 9.00 \\
S\&P100-Eq & 25.22 (25.12)& 1.98& 3.02& 14.72& \underline{76.36} & 10.66 \\
Naive & 17.89 (17.45)& 1.47& 2.14& \underline{14.71}& 65.09& 11.00 \\
MS-L-Cap & \textbf{66.22 (65.25)} & \textbf{2.90} & \textbf{4.95} & 22.81& 65.88& 9.64 \\
S\&P100 & 43.27 (43.20)& 2.86& 4.61& \textbf{16.17} & N/A & \textbf{9.24} \\
\hline
\end{tabular}
\begin{tablenotes}
\small
      \item[a]{Note: Values at bold indicate the best scores among the weighted strategies, while underlined values indicate the best scores among the equally weighted strategies. Values in parenthesis represent the total returns after transaction costs (5bps/trade).}
\end{tablenotes}
\end{table*}

\subsubsection{Rank-based strategies }
\begin{figure}[ht]
    \centering
        \includegraphics[width=\linewidth]{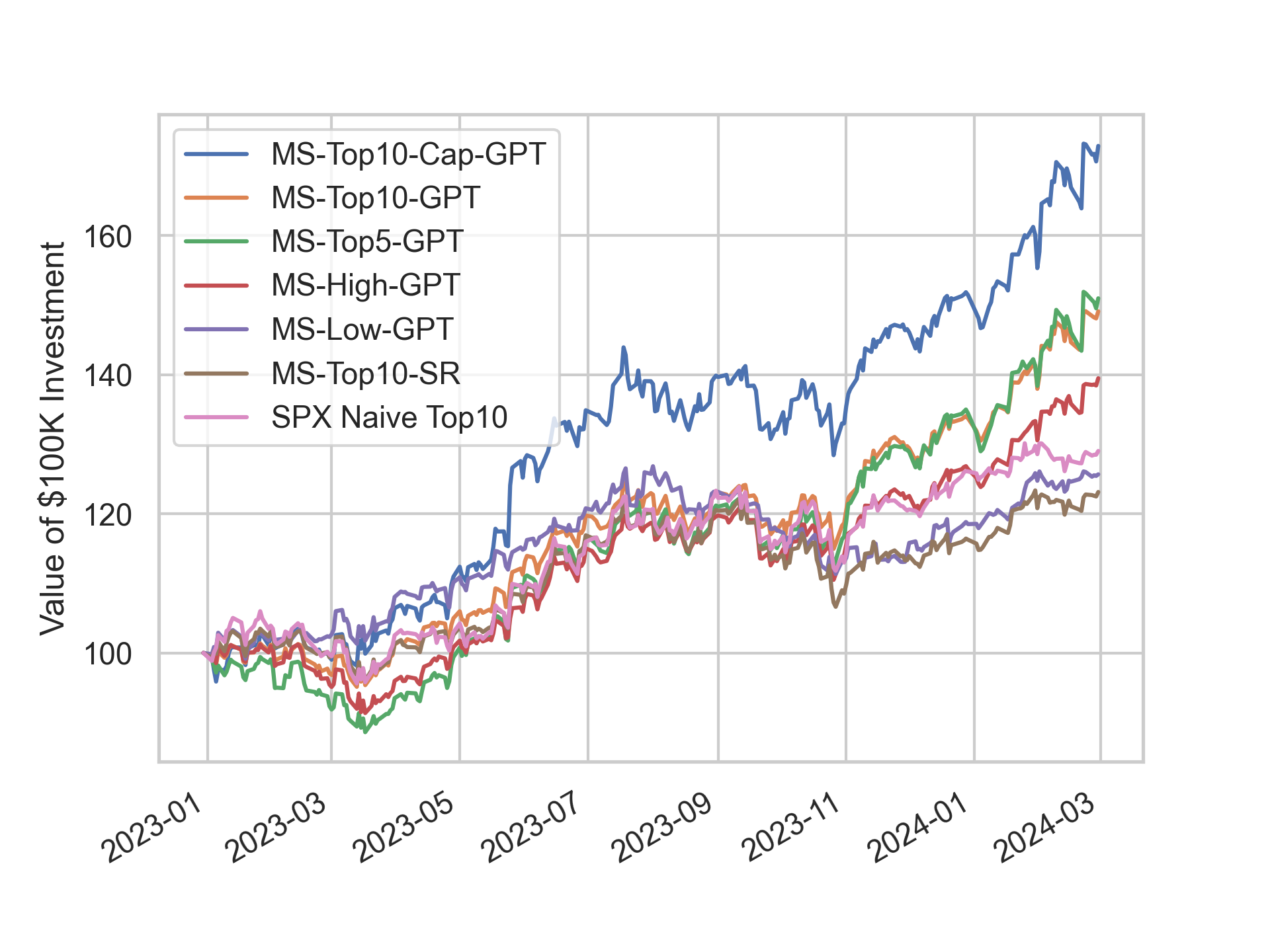}
        \caption{Performance of Ranked Portfolios}
        \label{fig:res3}
\end{figure}

Continuing the exploration of MarketSenseAI’s performance, Table \ref{tab:res2} and Figure \ref{fig:res3} provide a detailed analysis of rank-based  strategies derived from MarketSenseAI's signals. These results delve into the practical applications of these signals, focusing on portfolios with a manageable number of assets (around 10), and highlight how different strategic implementations can impact investment results.

\begin{table*}[ht]
\centering
\caption{MarketSenseAI Performance of Rank-Based Strategies} 
\label{tab:res2}
\begin{tabular}{lcccccc}
\hline
\textbf{Strategy} & \textbf{Total Return(\%)} & \textbf{Sharpe} & \textbf{Sortino} & \textbf{Vol(\%)}  & \textbf{Win Rate(\%)} & \textbf{Max Ddn(\%)} \\
\hline
MS-Top10-SR & 23.13 (22.12) & 1.45& 2.11& 19.27& 67.8& 12.66 \\
MS-Top5-GPT & 50.96 (49.67) & 2.26 & 3.69 & 24.01    & 68.42   & 11.39 \\
MS-Top10-GPT    & 49.09 (48.07)    & 2.68    & 4.29    & 19.37    & \textbf{74.1} & \textbf{7.66} \\
MS-High-GPT     & 39.47 (38.35)    & 2.28    & 3.44    & 19.08    & 71.9    & 9.73  \\
MS-Low-GPT& 25.66 (24.27)    & 1.76    & 2.64    & \textbf{17.04} & 55.1    & 12.22 \\
Naive-Top10     & 29.01 (28.18)    & 1.67    & 2.48    & 20.29    & 69.3    & 9.79 \\
MS-Top10-Cap-GPT& \textbf{72.87 (71.64)}& \textbf{2.80} & \textbf{4.89} & 25.61     &71.2     & 10.77\\
\hline
\end{tabular}
\begin{tablenotes}
\small
      \item[a]{Note: Values at bold indicate the best scores among strategies. Values in parenthesis represent the total returns after transaction costs (5bps/trade).}
\end{tablenotes}
\end{table*}

The MS-Top10-SR strategy, which selects the "buy" stocks with the highest Sharpe ratio, delivered a total return of 23.13\% (22.12\% after transaction costs). However, strategies utilizing GPT-4 to rank stocks exhibit significantly superior performance, both in total and risk-adjusted returns, highlighting the advanced analytical capabilities of LLM in selecting high-potential investments as well as the importance of the information generated by MarketSenseAI.

The MS-Top5-GPT and MS-Top10-GPT strategies, which select the top 5 and top 10 GPT-4 ranked stocks respectively, achieved remarkable total returns of 50.96\% (49.67\% after costs) and 49.09\% (48.07\% after costs). These strategies not only surpassed the returns of MS-Top10-SR and the Naive-Top10 benchmarks but also demonstrated impressive Sharpe and Sortino ratios, indicating high returns per unit of risk.

Notably, the MS-Top10-GPT strategy also achieved the highest win rate of 74.1\% and the lowest maximum drawdown of 7.66, further evidencing the robustness of GPT-4’s ranking mechanism in navigating market volatilities and capitalizing on growth opportunities.

The differentiation between MS-High-GPT and MS-Low-GPT strategies, yielding 39.47\% (38.35\% after costs) and 25.66\% (24.27\% after costs) respectively, elucidates the significant advantage of selecting stocks based on MarketSenseAI's explanations. This distinction highlights the added value of AI ranking over conventional financial engineering methods that may focus solely on historical returns or volatility.

Furthermore, the MS-Top10-Cap-GPT strategy, which applies a capitalization-weighted approach to the top GPT-4 ranked stocks, stands out with a staggering total return of 72.87\% (71.64\% after costs), the highest among the strategies evaluated.

\subsubsection{GPT ranking }
\begin{figure}[ht]
    \centering
        \includegraphics[width=\linewidth]{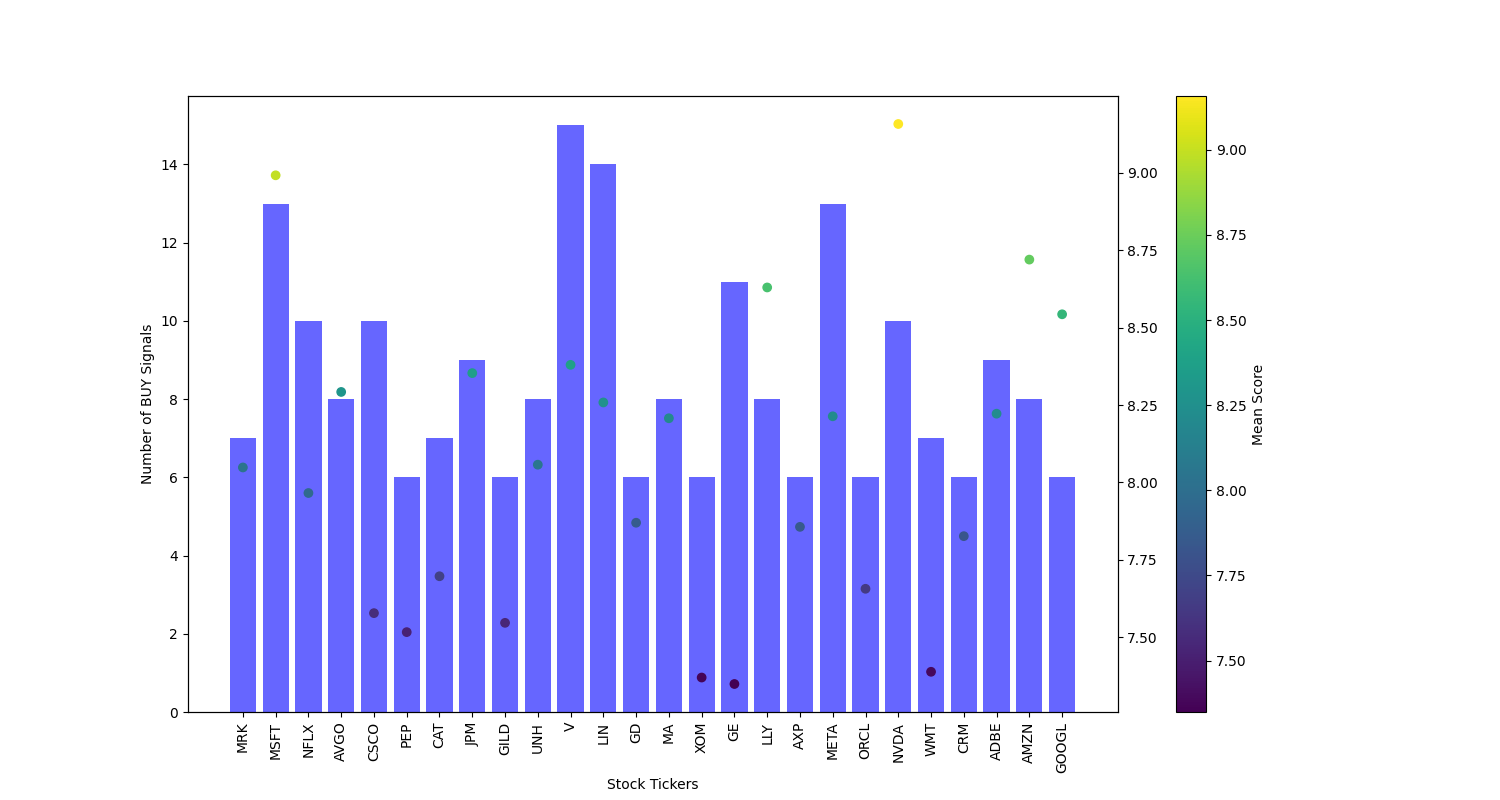}
        \caption{Signals Ranking by GPT-4}
        \label{fig:scoring}
\end{figure}

Building upon the insights gained from GPT-4’s rankings of MarketSenseAI signals, Figure~\ref{fig:scoring} provides a visual representation of the quality of explanations accompanying MarketSenseAI's "buy" signals throughout the evaluation period. The figure depicts (bar plot) the frequency of "buy" signals over the months for the stock with at least five "buy" signals. It also illustrates (scatter plot) the evaluative scores assigned by GPT-4, based on the strength, depth and relevance of explanations behind each "buy" recommendation. In this visualization, the scatter plot points are placed according to the average score given to each stock, allowing for an intuitive understanding of how the explanations for each stock were perceived in terms of quality and persuasiveness by the GPT-4 ranking layer.
 
A key observation from Figure~\ref{fig:scoring} is the significantly higher scoring of technology and AI-related stocks, such as Nvidia, Microsoft, and Amazon. This trend is reflective of the market hype surrounding these sectors during the evaluation period. The tendency of these stocks to receive higher scores not only demonstrates the relevance of MarketSenseAI's "buy" signals but also underscores the model's capability to encapsulate and convey the prevailing market sentiment and potential in its explanations.
\section{Conclusion}
\label{sec:6}
MarketSenseAI's innovative use of GPT-4's advanced reasoning capabilities represents a significant leap forward in financial analysis. This AI-powered framework provides both retail and professional investors with a comprehensive and unique tool for processing and analyzing extensive data sets, enabling the generation of scalable, actionable investment signals. Its utility is particularly evident in single stock analysis within environments rich in untapped opportunities.

The presented LLM-based system provides several key advantages over traditional financial tools, enhancing qualitative and quantitative data processing, mitigating biases, explaining its reasoning, and boosting scalability and efficiency. However, it's crucial to acknowledge the study's limitations, including the assumptions made about investment decision factors and GPT-4's effectiveness in different markets. The limited duration of the study also highlights the need for extended research to fully understand MarketSenseAI's effectiveness across various market conditions.

MarketSenseAI's evaluation in the S\&P 100, yielding up to 73\% returns and surpassing market performance by 30\%, showcases its potential in investment decision-making. In addition, beyond the specific signals generated by MarketSenseAI, its individual components independently provide substantial value. They deliver a clear and concise picture of each factor affecting a stock, combining summarization capabilities with comprehensive data aggregation.

These results contribute to understanding the role of LLMs in finance, suggesting new possibilities for future advancements in this dynamic and evolving field. MarketSenseAI represents a step forward in the process of investment decision-making and shows potential in supporting the overall health and stability of the financial system.

\section*{CRediT authorship contribution statement}
\textbf{George Fatouros}: Conceptualization, Methodology, Software, Validation, Formal analysis, Investigation, Data curation, Writing – original draft \& Visualization. \textbf{Kostas Metaxas}: Conceptualization, Methodology, Software, Validation, Investigation, Data curation, Writing - Review \& Editing. \textbf{John Soldatos}: Funding acquisition, Writing – review \& editing. \textbf{Dimosthenis Kyriazis}: Resources \& Supervision.

\section*{Declaration of Competing Interest}
The authors George Fatouros and Kostas Metaxas declare their involvement in the development and commercialization of the presented framework (i.e., MarketSenseAI) as a product of Alpha Tensor Technologies Ltd.


\begin{thebibliography}{50}
\expandafter\ifx\csname natexlab\endcsname\relax\def\natexlab#1{#1}\fi
\providecommand{\url}[1]{\texttt{#1}}
\providecommand{\href}[2]{#2}
\providecommand{\path}[1]{#1}
\providecommand{\DOIprefix}{doi:}
\providecommand{\ArXivprefix}{arXiv:}
\providecommand{\URLprefix}{URL: }
\providecommand{\Pubmedprefix}{pmid:}
\providecommand{\doi}[1]{\href{http://dx.doi.org/#1}{\path{#1}}}
\providecommand{\Pubmed}[1]{\href{pmid:#1}{\path{#1}}}
\providecommand{\bibinfo}[2]{#2}
\ifx\xfnm\relax \def\xfnm[#1]{\unskip,\space#1}\fi
\bibitem[{Abramski et~al.(2023)Abramski, Citraro, Lombardi, Rossetti and Stella}]{abramski2023cognitive}
\bibinfo{author}{Abramski, K.}, \bibinfo{author}{Citraro, S.}, \bibinfo{author}{Lombardi, L.}, \bibinfo{author}{Rossetti, G.}, \bibinfo{author}{Stella, M.}, \bibinfo{year}{2023}.
\newblock \bibinfo{title}{Cognitive network science reveals bias in gpt-3, gpt-3.5 turbo, and gpt-4 mirroring math anxiety in high-school students}.
\newblock \bibinfo{journal}{Big Data and Cognitive Computing} \bibinfo{volume}{7}, \bibinfo{pages}{124}.
\bibitem[{Alqahtani et~al.(2020)Alqahtani, Wither, Dong and Goodwin}]{alqahtani2020impact}
\bibinfo{author}{Alqahtani, A.}, \bibinfo{author}{Wither, M.J.}, \bibinfo{author}{Dong, Z.}, \bibinfo{author}{Goodwin, K.R.}, \bibinfo{year}{2020}.
\newblock \bibinfo{title}{Impact of news-based equity market volatility on international stock markets}.
\newblock \bibinfo{journal}{Journal of Applied Economics} \bibinfo{volume}{23}, \bibinfo{pages}{224--234}.
\bibitem[{Alshami et~al.(2023)Alshami, Elsayed, Ali, Eltoukhy and Zayed}]{alshami2023harnessing}
\bibinfo{author}{Alshami, A.}, \bibinfo{author}{Elsayed, M.}, \bibinfo{author}{Ali, E.}, \bibinfo{author}{Eltoukhy, A.E.}, \bibinfo{author}{Zayed, T.}, \bibinfo{year}{2023}.
\newblock \bibinfo{title}{Harnessing the power of chatgpt for automating systematic review process: Methodology, case study, limitations, and future directions}.
\newblock \bibinfo{journal}{Systems} \bibinfo{volume}{11}, \bibinfo{pages}{351}.
\bibitem[{Anand and Pathak(2022)}]{anand2022role}
\bibinfo{author}{Anand, A.}, \bibinfo{author}{Pathak, J.}, \bibinfo{year}{2022}.
\newblock \bibinfo{title}{The role of reddit in the gamestop short squeeze}.
\newblock \bibinfo{journal}{Economics Letters} \bibinfo{volume}{211}, \bibinfo{pages}{110249}.
\bibitem[{Araci(2019)}]{araci2019finbert}
\bibinfo{author}{Araci, D.}, \bibinfo{year}{2019}.
\newblock \bibinfo{title}{Finbert: Financial sentiment analysis with pre-trained language models}.
\newblock \bibinfo{note}{Preprint at \url{https://arxiv.org/abs/1908.10063}}.
\bibitem[{Atreides and Kelley(2023)}]{atreidescognitive}
\bibinfo{author}{Atreides, K.}, \bibinfo{author}{Kelley, D.}, \bibinfo{year}{2023}.
\newblock \bibinfo{title}{Cognitive biases in natural language: Automatically detecting, differentiating, and measuring bias in text}.
\newblock \DOIprefix\doi{10.13140/RG.2.2.14044.56967}.
\bibitem[{BIS(2022)}]{bis-central-bank-tools}
\bibinfo{author}{BIS}, \bibinfo{year}{2022}.
\newblock \bibinfo{title}{Market dysfunction and central bank tools}.
\newblock \bibinfo{note}{\url{https://www.bis.org/publ/mc\_insights.pdf.} Accessed September 28, 2023}.
\bibitem[{Bloomberg(2019)}]{bloomberg2019algo}
\bibinfo{author}{Bloomberg}, \bibinfo{year}{2019}.
\newblock \bibinfo{title}{What’s an “algo wheel?” and why should you care? | bloomberg professional services}.
\newblock \bibinfo{note}{\url{https://www.bloomberg.com/professional/blog/whats-algo-wheel-care/}. Accessed September 24, 2023}.
\bibitem[{Bouchaud et~al.(2003)Bouchaud, Gefen, Potters and Wyart}]{bouchaud2003fluctuations}
\bibinfo{author}{Bouchaud, J.P.}, \bibinfo{author}{Gefen, Y.}, \bibinfo{author}{Potters, M.}, \bibinfo{author}{Wyart, M.}, \bibinfo{year}{2003}.
\newblock \bibinfo{title}{Fluctuations and response in financial markets: the subtle nature ofrandom'price changes}.
\newblock \bibinfo{journal}{Quantitative finance} \bibinfo{volume}{4}, \bibinfo{pages}{176}.
\bibitem[{Brogaard et~al.(2023)Brogaard, Han and Won}]{brogaard2023does}
\bibinfo{author}{Brogaard, J.}, \bibinfo{author}{Han, J.}, \bibinfo{author}{Won, P.Y.}, \bibinfo{year}{2023}.
\newblock \bibinfo{title}{How does zero-day-to-expiry options trading affect the volatility of underlying assets?}
\newblock \bibinfo{note}{Available at SSRN: \url{https://ssrn.com/abstract=4426358} or \url{http://dx.doi.org/10.2139/ssrn.4426358}}.
\bibitem[{Callanan et~al.(2023)Callanan, Mbakwe, Papadimitriou, Pei, Sibue, Zhu, Ma, Liu and Shah}]{callanan2023can}
\bibinfo{author}{Callanan, E.}, \bibinfo{author}{Mbakwe, A.}, \bibinfo{author}{Papadimitriou, A.}, \bibinfo{author}{Pei, Y.}, \bibinfo{author}{Sibue, M.}, \bibinfo{author}{Zhu, X.}, \bibinfo{author}{Ma, Z.}, \bibinfo{author}{Liu, X.}, \bibinfo{author}{Shah, S.}, \bibinfo{year}{2023}.
\newblock \bibinfo{title}{Can gpt models be financial analysts? an evaluation of chatgpt and gpt-4 on mock cfa exams}.
\newblock \bibinfo{note}{Preprint at \url{https://arxiv.org/abs/2310.08678}}.
\bibitem[{Chase(2022)}]{lang2022}
\bibinfo{author}{Chase, H.}, \bibinfo{year}{2022}.
\newblock \bibinfo{title}{Langchain}.
\newblock \bibinfo{note}{\url{https://github.com/langchain-ai/langchain}. Accessed December 29, 2023}.
\bibitem[{Chen et~al.(2023)Chen, Zheng, Lu, Yuan and Zhu}]{chen2023chatgpt}
\bibinfo{author}{Chen, Z.}, \bibinfo{author}{Zheng, L.N.}, \bibinfo{author}{Lu, C.}, \bibinfo{author}{Yuan, J.}, \bibinfo{author}{Zhu, D.}, \bibinfo{year}{2023}.
\newblock \bibinfo{title}{Chatgpt informed graph neural network for stock movement prediction}.
\newblock \bibinfo{note}{Preprint at \url{https://arxiv.org/abs/2306.03763}}.
\bibitem[{Chui et~al.(2023)Chui, Hazan, Roberts, Singla, Smaje, Sukharevsky, Yee and Zemmel}]{mckinsey2023}
\bibinfo{author}{Chui, M.}, \bibinfo{author}{Hazan, E.}, \bibinfo{author}{Roberts, R.}, \bibinfo{author}{Singla, A.}, \bibinfo{author}{Smaje, K.}, \bibinfo{author}{Sukharevsky, A.}, \bibinfo{author}{Yee, L.}, \bibinfo{author}{Zemmel, R.}, \bibinfo{year}{2023}.
\newblock \bibinfo{title}{The Economic Potential of Generative AI: The Next Productivity Frontier}.
\newblock \bibinfo{type}{Technical Report}. McKinsey \& Company.
\newblock \bibinfo{note}{\url{https://www.mckinsey.com/capabilities/mckinsey-digital/our-insights/the-economic-potential-of-generative-ai-the-next-productivity-frontier}. Accessed September 24, 2023}.
\bibitem[{CNBC(2023)}]{jpmorgan-gpt}
\bibinfo{author}{CNBC}, \bibinfo{year}{2023}.
\newblock \bibinfo{title}{Jpmorgan ai investment advisor}.
\newblock \bibinfo{note}{\url{https://www.cnbc.com/2023/05/25/jpmorgan-develops-ai-investment-advisor.html}. Accessed September 24, 2023}.
\bibitem[{Dong et~al.(2022)Dong, Li, Dai, Zheng, Wu, Chang, Sun, Xu and Sui}]{dong2022survey}
\bibinfo{author}{Dong, Q.}, \bibinfo{author}{Li, L.}, \bibinfo{author}{Dai, D.}, \bibinfo{author}{Zheng, C.}, \bibinfo{author}{Wu, Z.}, \bibinfo{author}{Chang, B.}, \bibinfo{author}{Sun, X.}, \bibinfo{author}{Xu, J.}, \bibinfo{author}{Sui, Z.}, \bibinfo{year}{2022}.
\newblock \bibinfo{title}{A survey for in-context learning}.
\newblock \bibinfo{note}{Preprint at \url{https://arxiv.org/abs/2301.00234}}.
\bibitem[{Efron and Tibshirani(1986)}]{efron1986bootstrap}
\bibinfo{author}{Efron, B.}, \bibinfo{author}{Tibshirani, R.}, \bibinfo{year}{1986}.
\newblock \bibinfo{title}{Bootstrap methods for standard errors, confidence intervals, and other measures of statistical accuracy}.
\newblock \bibinfo{journal}{Statistical science} \bibinfo{volume}{1}, \bibinfo{pages}{54--75}.
\bibitem[{Fatouros et~al.(2023a)Fatouros, Makridis, Kotios, Soldatos, Filippakis and Kyriazis}]{fatouros2023deepvar}
\bibinfo{author}{Fatouros, G.}, \bibinfo{author}{Makridis, G.}, \bibinfo{author}{Kotios, D.}, \bibinfo{author}{Soldatos, J.}, \bibinfo{author}{Filippakis, M.}, \bibinfo{author}{Kyriazis, D.}, \bibinfo{year}{2023}a.
\newblock \bibinfo{title}{Deepvar: a framework for portfolio risk assessment leveraging probabilistic deep neural networks}.
\newblock \bibinfo{journal}{Digital finance} \bibinfo{volume}{5}, \bibinfo{pages}{29--56}.
\bibitem[{Fatouros et~al.(2023b)Fatouros, Soldatos, Kouroumali, Makridis and Kyriazis}]{fatouros2023transforming}
\bibinfo{author}{Fatouros, G.}, \bibinfo{author}{Soldatos, J.}, \bibinfo{author}{Kouroumali, K.}, \bibinfo{author}{Makridis, G.}, \bibinfo{author}{Kyriazis, D.}, \bibinfo{year}{2023}b.
\newblock \bibinfo{title}{Transforming sentiment analysis in the financial domain with chatgpt}.
\newblock \bibinfo{journal}{Machine Learning with Applications} \bibinfo{volume}{14}, \bibinfo{pages}{100508}.
\bibitem[{Fontinelle(2022)}]{invest2022}
\bibinfo{author}{Fontinelle, A.}, \bibinfo{year}{2022}.
\newblock \bibinfo{title}{Can anybody beat the market?}
\newblock \bibinfo{note}{\url{https://www.investopedia.com/ask/answers/12/beating-the-market.asp}, Accessed January 01, 2024}.
\bibitem[{Goyal and He(2015)}]{goyal2015passive}
\bibinfo{author}{Goyal, A.}, \bibinfo{author}{He, Z.}, \bibinfo{year}{2015}.
\newblock \bibinfo{title}{Passive investing and market liquidity}.
\newblock \bibinfo{journal}{The Review of Financial Studies} \bibinfo{volume}{28}, \bibinfo{pages}{2167--2203}.
\bibitem[{Greenwald et~al.(2020)Greenwald, Kahn, Bellissimo, Cooper and Santos}]{greenwald2020value}
\bibinfo{author}{Greenwald, B.C.}, \bibinfo{author}{Kahn, J.}, \bibinfo{author}{Bellissimo, E.}, \bibinfo{author}{Cooper, M.A.}, \bibinfo{author}{Santos, T.}, \bibinfo{year}{2020}.
\newblock \bibinfo{title}{Value investing: from Graham to Buffett and beyond}.
\newblock \bibinfo{publisher}{John Wiley \& Sons}.
\bibitem[{Guo et~al.(2023)Guo, Zhang, Wang, Jiang, Nie, Ding, Yue and Wu}]{guo2023close}
\bibinfo{author}{Guo, B.}, \bibinfo{author}{Zhang, X.}, \bibinfo{author}{Wang, Z.}, \bibinfo{author}{Jiang, M.}, \bibinfo{author}{Nie, J.}, \bibinfo{author}{Ding, Y.}, \bibinfo{author}{Yue, J.}, \bibinfo{author}{Wu, Y.}, \bibinfo{year}{2023}.
\newblock \bibinfo{title}{How close is chatgpt to human experts? comparison corpus, evaluation, and detection}.
\newblock \bibinfo{note}{Preprint at \url{https://arxiv.org/abs/2301.07597}}.
\bibitem[{Kidwell et~al.(2016)Kidwell, Blackwell and Whidbee}]{kidwell2016financial}
\bibinfo{author}{Kidwell, D.S.}, \bibinfo{author}{Blackwell, D.W.}, \bibinfo{author}{Whidbee, D.A.}, \bibinfo{year}{2016}.
\newblock \bibinfo{title}{Financial institutions, markets, and money}.
\newblock \bibinfo{publisher}{John Wiley \& Sons}.
\bibitem[{Kim et~al.(2023)Kim, Muhn and Nikolaev}]{kim2023bloated}
\bibinfo{author}{Kim, A.G.}, \bibinfo{author}{Muhn, M.}, \bibinfo{author}{Nikolaev, V.V.}, \bibinfo{year}{2023}.
\newblock \bibinfo{title}{Bloated disclosures: Can chatgpt help investors process information?}
\newblock \bibinfo{note}{Available at SSRN: \url{https://ssrn.com/abstract=4425527} or \url{http://dx.doi.org/10.2139/ssrn.4425527}}.
\bibitem[{Kirtac and Germano(2024)}]{kirtac2024sentiment}
\bibinfo{author}{Kirtac, K.}, \bibinfo{author}{Germano, G.}, \bibinfo{year}{2024}.
\newblock \bibinfo{title}{Sentiment trading with large language models}.
\newblock \bibinfo{note}{Available at SSRN: \url{https://ssrn.com/abstract=4706629}}.
\bibitem[{Korn et~al.(2022)Korn, M{\"o}ller and Schwehm}]{korn2022drawdown}
\bibinfo{author}{Korn, O.}, \bibinfo{author}{M{\"o}ller, P.M.}, \bibinfo{author}{Schwehm, C.}, \bibinfo{year}{2022}.
\newblock \bibinfo{title}{Drawdown measures: Are they all the same?}
\newblock \bibinfo{journal}{The Journal of Portfolio Management} \bibinfo{volume}{48}, \bibinfo{pages}{104--120}.
\bibitem[{Kotios et~al.(2022)Kotios, Makridis, Fatouros and Kyriazis}]{kotios2022deep}
\bibinfo{author}{Kotios, D.}, \bibinfo{author}{Makridis, G.}, \bibinfo{author}{Fatouros, G.}, \bibinfo{author}{Kyriazis, D.}, \bibinfo{year}{2022}.
\newblock \bibinfo{title}{Deep learning enhancing banking services: a hybrid transaction classification and cash flow prediction approach}.
\newblock \bibinfo{journal}{Journal of big Data} \bibinfo{volume}{9}, \bibinfo{pages}{100}.
\bibitem[{Lewellen(2002)}]{lewellen2002momentum}
\bibinfo{author}{Lewellen, J.}, \bibinfo{year}{2002}.
\newblock \bibinfo{title}{Momentum and autocorrelation in stock returns}.
\newblock \bibinfo{journal}{The Review of Financial Studies} \bibinfo{volume}{15}, \bibinfo{pages}{533--564}.
\bibitem[{Li et~al.(2023)Li, Zhu, Ma, Liu and Shah}]{li2023chatgpt}
\bibinfo{author}{Li, X.}, \bibinfo{author}{Zhu, X.}, \bibinfo{author}{Ma, Z.}, \bibinfo{author}{Liu, X.}, \bibinfo{author}{Shah, S.}, \bibinfo{year}{2023}.
\newblock \bibinfo{title}{Are chatgpt and gpt-4 general-purpose solvers for financial text analytics? an examination on several typical tasks}.
\newblock \bibinfo{note}{Preprint at \url{https://arxiv.org/abs/2305.05862}}.
\bibitem[{Liu et~al.(2023)Liu, Iter, Xu, Wang, Xu and Zhu}]{liu2023gpteval}
\bibinfo{author}{Liu, Y.}, \bibinfo{author}{Iter, D.}, \bibinfo{author}{Xu, Y.}, \bibinfo{author}{Wang, S.}, \bibinfo{author}{Xu, R.}, \bibinfo{author}{Zhu, C.}, \bibinfo{year}{2023}.
\newblock \bibinfo{title}{Gpteval: Nlg evaluation using gpt-4 with better human alignment}.
\newblock \bibinfo{note}{Preprint at \url{https://arxiv.org/abs/2303.16634}}.
\bibitem[{Lopez-Lira and Tang(2023)}]{lopez2023can}
\bibinfo{author}{Lopez-Lira, A.}, \bibinfo{author}{Tang, Y.}, \bibinfo{year}{2023}.
\newblock \bibinfo{title}{Can chatgpt forecast stock price movements? return predictability and large language models}.
\newblock \bibinfo{note}{Preprint at \url{https://arxiv.org/abs/2304.07619}}.
\bibitem[{LTXtrading(2023)}]{bondgpt}
\bibinfo{author}{LTXtrading}, \bibinfo{year}{2023}.
\newblock \bibinfo{title}{Bondgpt: Introducing ltx’s generative ai application for corporate bond trading.}
\newblock \bibinfo{note}{\url{https://www.ltxtrading.com/bondgpt} Accessed January 19, 2024}.
\bibitem[{Malik(2011)}]{malik2011estimating}
\bibinfo{author}{Malik, F.}, \bibinfo{year}{2011}.
\newblock \bibinfo{title}{Estimating the impact of good news on stock market volatility}.
\newblock \bibinfo{journal}{Applied Financial Economics} \bibinfo{volume}{21}, \bibinfo{pages}{545--554}.
\bibitem[{Malkiel(2003)}]{malkiel2003efficient}
\bibinfo{author}{Malkiel, B.G.}, \bibinfo{year}{2003}.
\newblock \bibinfo{title}{The efficient market hypothesis and its critics}.
\newblock \bibinfo{journal}{Journal of economic perspectives} \bibinfo{volume}{17}, \bibinfo{pages}{59--82}.
\bibitem[{Metaxas(2023)}]{marketdigest-cube}
\bibinfo{author}{Metaxas, K.}, \bibinfo{year}{2023}.
\newblock \bibinfo{title}{Marketdigest}.
\newblock \bibinfo{note}{\url{https://www.km3am.com/2023/03/13/marketdigest-new-ai-powered-tool-for-wealth-management-insights/}. Accessed September 24, 2023}.
\bibitem[{Noy and Zhang(2023)}]{noy2023experimental}
\bibinfo{author}{Noy, S.}, \bibinfo{author}{Zhang, W.}, \bibinfo{year}{2023}.
\newblock \bibinfo{title}{Experimental evidence on the productivity effects of generative artificial intelligence}.
\newblock \bibinfo{journal}{Science} \bibinfo{volume}{381}, \bibinfo{pages}{187--192}.
\newblock \URLprefix \url{https://www.science.org/doi/abs/10.1126/science.adh2586}, \DOIprefix\doi{10.1126/science.adh2586}.
\bibitem[{OECD(2021)}]{oecd2021}
\bibinfo{author}{OECD}, \bibinfo{year}{2021}.
\newblock \bibinfo{title}{Artificial intelligence, machine learning and big data in finance: Opportunities, challenges, and implications for policy makers}.
\newblock \bibinfo{note}{\url{https://www.oecd.org/finance/financial-markets/Artificial-intelligence-machine-learning-big-data-in-finance.pdf}. Accessed September 24, 2023}.
\bibitem[{OpenAI(2023a)}]{openai2023gpt4}
\bibinfo{author}{OpenAI}, \bibinfo{year}{2023}a.
\newblock \bibinfo{title}{Gpt-4 technical report}.
\newblock \href{http://arxiv.org/abs/2303.08774}{{\tt arXiv:2303.08774}}. \bibinfo{note}{preprint at \url{https://arxiv.org/abs/2303.08774}}.
\bibitem[{OpenAI(2023b)}]{ms-openai}
\bibinfo{author}{OpenAI}, \bibinfo{year}{2023}b.
\newblock \bibinfo{title}{Morgan stanley wealth management deploys gpt-4 to organize its vast knowledge base.}
\newblock \bibinfo{note}{\url{https://openai.com/customer-stories/morgan-stanley}. Accessed January 19, 2024}.
\bibitem[{Shu et~al.(2023)Shu, Wichers, Luo, Zhu, Liu, Chen and Meng}]{shu2023fusion}
\bibinfo{author}{Shu, L.}, \bibinfo{author}{Wichers, N.}, \bibinfo{author}{Luo, L.}, \bibinfo{author}{Zhu, Y.}, \bibinfo{author}{Liu, Y.}, \bibinfo{author}{Chen, J.}, \bibinfo{author}{Meng, L.}, \bibinfo{year}{2023}.
\newblock \bibinfo{title}{Fusion-eval: Integrating evaluators with llms}.
\newblock \bibinfo{note}{Preprint at \url{https://arxiv.org/abs/2311.09204}}.
\bibitem[{Song et~al.(2020)Song, Tan, Qin, Lu and Liu}]{song2020mpnet}
\bibinfo{author}{Song, K.}, \bibinfo{author}{Tan, X.}, \bibinfo{author}{Qin, T.}, \bibinfo{author}{Lu, J.}, \bibinfo{author}{Liu, T.Y.}, \bibinfo{year}{2020}.
\newblock \bibinfo{title}{Mpnet: Masked and permuted pre-training for language understanding}.
\newblock \bibinfo{journal}{Advances in Neural Information Processing Systems} \bibinfo{volume}{33}, \bibinfo{pages}{16857--16867}.
\bibitem[{Tetlock et~al.(2008)Tetlock, Saar-Tsechansky and Macskassy}]{tetlock2008more}
\bibinfo{author}{Tetlock, P.C.}, \bibinfo{author}{Saar-Tsechansky, M.}, \bibinfo{author}{Macskassy, S.}, \bibinfo{year}{2008}.
\newblock \bibinfo{title}{More than words: Quantifying language to measure firms' fundamentals}.
\newblock \bibinfo{journal}{The journal of finance} \bibinfo{volume}{63}, \bibinfo{pages}{1437--1467}.
\bibitem[{Thompson(2023)}]{magnificent2023}
\bibinfo{author}{Thompson, C.}, \bibinfo{year}{2023}.
\newblock \bibinfo{title}{Magnificent 7 stocks: What you need to know}.
\newblock \bibinfo{note}{\url{https://www.investopedia.com/magnificent-seven-stocks-8402262}, Accessed January 01, 2024}.
\bibitem[{Tjuatja et~al.(2023)Tjuatja, Chen, Wu, Talwalkar and Neubig}]{tjuatja2023llms}
\bibinfo{author}{Tjuatja, L.}, \bibinfo{author}{Chen, V.}, \bibinfo{author}{Wu, S.T.}, \bibinfo{author}{Talwalkar, A.}, \bibinfo{author}{Neubig, G.}, \bibinfo{year}{2023}.
\newblock \bibinfo{title}{Do llms exhibit human-like response biases? a case study in survey design}.
\newblock \bibinfo{note}{Preprint at \url{https://arxiv.org/abs/2311.04076}}.
\bibitem[{Wei et~al.(2022)Wei, Wang, Schuurmans, Bosma, Xia, Chi, Le, Zhou et~al.}]{wei2022chain}
\bibinfo{author}{Wei, J.}, \bibinfo{author}{Wang, X.}, \bibinfo{author}{Schuurmans, D.}, \bibinfo{author}{Bosma, M.}, \bibinfo{author}{Xia, F.}, \bibinfo{author}{Chi, E.}, \bibinfo{author}{Le, Q.V.}, \bibinfo{author}{Zhou, D.}, et~al., \bibinfo{year}{2022}.
\newblock \bibinfo{title}{Chain-of-thought prompting elicits reasoning in large language models}.
\newblock \bibinfo{journal}{Advances in Neural Information Processing Systems} \bibinfo{volume}{35}, \bibinfo{pages}{24824--24837}.
\bibitem[{Weiss-Cohen et~al.(2019)Weiss-Cohen, Ayton, Clacher and Thoma}]{weiss2019behavioral}
\bibinfo{author}{Weiss-Cohen, L.}, \bibinfo{author}{Ayton, P.}, \bibinfo{author}{Clacher, I.}, \bibinfo{author}{Thoma, V.}, \bibinfo{year}{2019}.
\newblock \bibinfo{title}{Behavioral biases in pension fund trustees’ decision making}.
\newblock \bibinfo{journal}{Review of Behavioral Finance} \bibinfo{volume}{11}, \bibinfo{pages}{128--143}.
\bibitem[{Wu et~al.(2023)Wu, Irsoy, Lu, Dabravolski, Dredze, Gehrmann, Kambadur, Rosenberg and Mann}]{bloomberg-gpt}
\bibinfo{author}{Wu, S.}, \bibinfo{author}{Irsoy, O.}, \bibinfo{author}{Lu, S.}, \bibinfo{author}{Dabravolski, V.}, \bibinfo{author}{Dredze, M.}, \bibinfo{author}{Gehrmann, S.}, \bibinfo{author}{Kambadur, P.}, \bibinfo{author}{Rosenberg, D.}, \bibinfo{author}{Mann, G.}, \bibinfo{year}{2023}.
\newblock \bibinfo{title}{Bloomberggpt: A large language model for finance}.
\newblock \bibinfo{note}{Preprint at \url{https://arxiv.org/abs/2303.17564}}.
\bibitem[{Yu et~al.(2023)Yu, Chen, Ling, Dong, Liu and Lu}]{yu2023temporal}
\bibinfo{author}{Yu, X.}, \bibinfo{author}{Chen, Z.}, \bibinfo{author}{Ling, Y.}, \bibinfo{author}{Dong, S.}, \bibinfo{author}{Liu, Z.}, \bibinfo{author}{Lu, Y.}, \bibinfo{year}{2023}.
\newblock \bibinfo{title}{Temporal data meets llm--explainable financial time series forecasting}.
\newblock \bibinfo{note}{Preprint at \url{https://arxiv.org/abs/2306.11025}}.
\bibitem[{Zaremba and Demir(2023)}]{zaremba2023chatgpt}
\bibinfo{author}{Zaremba, A.}, \bibinfo{author}{Demir, E.}, \bibinfo{year}{2023}.
\newblock \bibinfo{title}{Chatgpt: Unlocking the future of nlp in finance}.
\newblock \bibinfo{note}{Available at SSRN: \url{https://ssrn.com/abstract=4323643} or h\url{ttp://dx.doi.org/10.2139/ssrn.4323643}}.

\end{thebibliography}


\end{document}